\documentclass[
 twocolumn,
 amsmath,amssymb, 
 aps,
 prd,
 letterpaper,
 superscriptaddress,
 showpacs, showkeys,
floatfix,
 nofootinbib,
]{revtex4-2}% 

\usepackage[colorlinks,linkcolor=blue,anchorcolor=blue, citecolor=blue,urlcolor=blue,]{hyperref}
\usepackage{graphicx}
\usepackage{color}
\usepackage{xcolor}
\usepackage{float}
\usepackage{makecell}
\usepackage{newtxtext,newtxmath}
\usepackage{fancyhdr}
\usepackage{hyperref}
\usepackage{xcolor}
\allowdisplaybreaks
\usepackage{graphicx}
\usepackage{appendix}
\usepackage{amssymb}
\usepackage{multirow}
\usepackage{booktabs}
\usepackage{float} 
\usepackage{subfigure}
\usepackage{natbib}
\usepackage[marginal]{footmisc}
\usepackage{tikz}
\usepackage{tabularx}
\usepackage{algorithm}
\usepackage{algorithmic}
\usepackage{xcolor}
\usepackage{aas_macros}
\usepackage{enumitem}
\usepackage{ulem}
\usepackage{footnote}

\makeatletter
\usetikzlibrary{shapes.geometric, arrows.meta, positioning}
\newcommand{\arcmin}{^\prime}

\newcommand{\akra}{{AKRA 3.0}}
\newcommand{\ellmax}{\ell_{\mathrm{max}}}
\begin{document}

\title{AKRA 3.0: A matrix-free Inversion Framework for Weak Lensing Mass Mapping \\ and Its Application to DES Y3 Data }

\author{Yuan Shi}
\email{yshi@sjtu.edu.cn}
\affiliation{State Key Laboratory of Dark Matter Physics, School of Physics and Astronomy, Shanghai Jiao Tong University, Shanghai 200240, China}
\affiliation{Key Laboratory for Particle Astrophysics and Cosmology (MOE) / Shanghai Key Laboratory for Particle Physics and Cosmology, China}

\author{Pengjie Zhang}
\email{zhangpj@sjtu.edu.cn} 
\affiliation{State Key Laboratory of Dark Matter Physics, School of Physics and Astronomy, Shanghai Jiao Tong University, Shanghai 200240, China}
\affiliation{Key Laboratory for Particle Astrophysics and Cosmology (MOE) / Shanghai Key Laboratory for Particle Physics and Cosmology, China}
\affiliation{Division of Astronomy and Astrophysics, Tsung-Dao Lee Institute, Shanghai Jiao Tong University, Shanghai, 200240, China}

\author{Li Cui}
\affiliation{State Key Laboratory of Dark Matter Physics, School of Physics and Astronomy, Shanghai Jiao Tong University, Shanghai 200240, China}
\affiliation{Key Laboratory for Particle Astrophysics and Cosmology (MOE) / Shanghai Key Laboratory for Particle Physics and Cosmology, China}

\author{Jian Qin}
\affiliation{Center for Astronomy and Space Sciences, China Three Gorges University, Yichang 443000, People’s Republic of China}
\affiliation{College of Mathematics and Physics, China Three Gorges University, Yichang 443000, People’s Republic of China}

\author{Ji Yao}
\affiliation{Shanghai Astronomical Observatory (SHAO), Nandan Road 80, Shanghai, China}

\date{\today}% It is always \today, today,
             %  but any date may be explicitly specified

\begin{abstract}
  Weak gravitational lensing mass mapping offers a direct probe of the matter distribution.
  Accurate reconstruction of mass maps from masked shear catalogs remains challenging due to survey boundaries and spatially varying noise. In AKRA 2.0, we addressed the mask problem on the curved sky by constructing and inverting the normal-equation matrix $\bf{H} \equiv \mathbf{A}^\mathrm{T}\mathbf{N}^{-1} \mathbf{A}$ explicitly, necessitating a split-scale strategy that reconstructed different angular scales independently to reach high resolution.
  Here we present AKRA 3.0, in which $\mathbf{H}$ is treated as a linear operator and the normal equations are solved by the conjugate gradient (CG) method. 
  This reformulation reduces the memory requirement from $\mathcal{O}(N^2)$ to $\mathcal{O}(N)$ and the inversion cost from $\mathcal{O}(N^3)$ to $\mathcal{O}(N_{\rm iter}N^{3/2}), N \sim \ell_{\rm{max}}^2$ for full-sky (SHT-based) operations.
  % and $\mathcal{O}(N \log \sqrt{N}), N=n^2$ for flat-sky (2D FFT-based) operations.
  % $N$ represents the total number of modes or pixels, scaling as $N \sim \ell_{\rm{max}}^2$ for spherical geometry and $N \sim n^2$ for flat-sky grids.
  Such optimizations render high-resolution full-sky reconstruction tractable for Stage~III and Stage~IV surveys.
  Applying AKRA 3.0 to the DES Y3 \texttt{METACALIBRATION} catalog (4,143 $\deg^2$ and $\sim 10^8$ source galaxies), we produce the highest-resolution convergence map of this dataset to date at HEALPix $N_{\rm{nside}}= 2048$ without imposing any prior assumptions. 
  % ($\theta_{\rm{pix}}\sim 1.72 \arcmin, N_{\rm pix} = 5\times 10^7$) 
  % Fitting an overall lensing amplitude to the convergence-map power spectra yields $\hat{A}_L = 0.96 \pm 0.06$ from the split bin $\kappa^i\kappa^j$ cross-spectrum over $10 < \ell < 4000$ and $\hat{A}_L = 0.99 \pm 0.04$ from the $\kappa\kappa$ auto-spectrum over $10 < \ell < 2000$, both consistent with the fiducial $\Lambda$CDM prediction ($A_L = 1$) and demonstrating that unbiased two-point measurements can be obtained directly from the reconstructed map.
  We extract the convergence power spectrum directly from the reconstructed map and demonstrate that unbiased two-point measurements can be obtained directly from the reconstructed map.
The reconstructed E-mode convergence map will be publicly released as data products to enable future studies of non-Gaussian statistics, higher-order moments, and cross-correlations with external datasets.

\end{abstract}

\maketitle

\section{Introduction} \label{sec:intro}
Weak gravitational lensing has emerged as one of the most powerful probes of cosmology \cite{Bartelmann2001,Kilbinger2015,Mandelbaum2018}. By directly tracing the total matter distribution, weak lensing offers unique constraints on the amplitude of matter fluctuations $\sigma_8$, the matter density $\Omega_{\rm m}$, and the dark energy equation of state $w$ \cite{Weinberg2013}. 
Current Stage-III experiments, such as the Dark Energy Survey (DES) \cite{DES2016,Gatti2021}, the Kilo-Degree Survey (KiDS) \cite{Kuijken2019}, and the Hyper Suprime-Cam Survey (HSC) \cite{Aihara2018} have successfully measured cosmic shear with total signal-to-noise ratios $\mathrm{S/N}$ $\sim 40$ \cite{Amon2022,Secco2022}, rising to $\mathrm{S/N}\sim 83$ in recent DES~Y6 analysis \cite{DESY6shear2026}.
The upcoming Stage IV missions, including Euclid \cite{Euclid2024}, the Vera C. Rubin Observatory's Legacy Survey of Space and Time (LSST) \cite{LSST2009, Ivezi2020}, the Roman Space Telescope \cite{Spergel2015}, and the Chinese Space Station Telescope (CSST; \cite{Gong2019, Zhan2021}) will push the total $\mathrm{S/N}$ to $\gtrsim 300$ \cite{Yao2024CSST}, nearly an order of magnitude above early Stage-III analyses and a factor of $\sim$4 beyond the latest DES~Y6 measurement.

While most weak-lensing cosmological analyses rely primarily on shear ($\gamma$) statistics using two-point correlation functions or power spectra, fully exploiting this precision requires transitioning to the convergence ($\kappa$) field. 
First, the convergence is a scalar field, it provides a natural basis for extracting non-Gaussian information through higher-order moments, such as one-point probability distribution functions (PDFs, \cite{Chang2018, Barthelemy2020}), peak counts \cite{Shan2018,Martinet2018,Liu2023}, skewness and kurtosis \cite{VanWaerbeke2013,Petri2015}, Minkowski functionals \cite{Kratochvil2012,Vicinanza2019,Zurcher2021}, wavelet scattering transforms \cite{Cheng2020,Cheng2021}, as well as recent frameworks like machine learning \cite{Gupta2018,Ribli2019a,Fluri2022,LuTianhuan2023} and field-level inference \cite{Zhou2024,Zeghal2025}. 
Second, in galaxy--galaxy lensing the tangential shear $\gamma_t(\theta)$ at a given angular scale is inherently nonlocal, it depends on the projected mass distribution interior to $\theta$, mixing contributions from smaller scales $\theta' < \theta$ and complicating the theoretical modeling of small-scale signals \cite{Baldauf2010,MacCrann2020,Park2021,Prat2022}. In contrast, the $\kappa$-galaxy cross-correlation is free of this issue, as $\kappa$ is a local scalar.
Moreover, a prior-free convergence map constitutes an independent cosmological probe in its own right: two-point statistics measured from the map can be directly compared with shear-based estimates, providing a powerful consistency check and an alternative path to parameter constraints. 
% Finally, as a projected scalar field, the convergence map readily enables cross-correlations with other tracers of large-scale structure, CMB lensing convergence, the thermal Sunyaev--Zel'dovich Compton-$y$ parameter, galaxy density fields, and X-ray surface brightness, breaking parameter degeneracies and providing powerful self-calibration of systematic effects \cite{Chang2023, Eifler2024}. 
Together, these advantages highlight the convergence field as a natural bridge between complex observational data and theoretical predictions,  enhancing our ability to constrain cosmological parameters in future surveys.

However, reconstructing the convergence from observed shear catalog is a non-trivial inverse problem. The seminal Kaiser–Squires (KS) inversion \citep{Kaiser1993} recovers $\kappa$ from $\gamma$ in Fourier space under the idealized assumptions of a flat sky, with complete and uniform coverage. In practice, survey boundaries, masked regions, and spatially varying noise violate these assumptions, causing the KS estimator to produce biased maps \cite{Jeffrey2021}. A rich landscape of methods has been developed to address these challenges: Wiener filtering \cite{Jeffrey2018}, sparsity-based regularization (GLIMPSE; \cite{Leonard2014,Price2019}), Bayesian forward modeling (KaRMMa; \cite{Pier2022}), and deep-learning approaches such as score-based diffusion models. While these methods improve map quality by imposing a prior on the convergence field, which may alter its statistical properties.
Accurate mass mapping remains challenging: how can we recover the convergence field from noisy and incomplete shear field without losing information \cite{tegmark_how_1997}?

To address this rigorously, we developed the Accurate Kappa Reconstruction Algorithm (AKRA; \cite{Shi2024akra}) on the flat sky and its curved-sky extension AKRA 2.0 \cite{Shi2025}, which provide the unbiased minimum-variance linear estimator of the convergence field, along with an iterative extension \cite{Shi2026} for the galaxy cluster regime. The framework has been validated on simulations and successfully applied to HSC Y1 data \cite{Shi2025HSC}, demonstrating its ability to produce high-accuracy prior-free convergence maps. 
The AKRA series provides a complete theoretical solution. However, applying AKRA 2.0 at high resolution on wide-field data remains computationally prohibitive.

AKRA 2.0 mitigates computational costs by splitting the problem into large-scale (full-sky) and small-scale (flat-sky) regimes, which are solved independently. This is a physically motivated workaround, and the two scales must be solved, stored, and joined separately.
Yet, this only reduces the local problem size without altering the fundamental computational scaling. Within each regime the algorithm still constructs the normal-equation matrix $\mathbf{H} \equiv \mathbf{A}^\mathrm{T}\mathbf{N}^{-1}\mathbf{A}$
explicitly and subsequently inverts it. 
While the decomposition reduces the size of each sub-problem, it does not alter the underlying dense-matrix scaling, leaving the method computationally infeasible for Stage IV survey resolutions.
% Although the sub-problems are individually smaller, the dense-matrix scaling is unchanged, and at the resolutions demanded by Stage IV surveys the approach remains infeasible.

AKRA 3.0 overcomes the storage and inversion bottlenecks by combining three key ideas. First, convolution in Fourier space is equivalent to pointwise multiplication in real space. In previous work, we generate dense mask transfer matrix column by column using delta function or Toeplitz structures. AKRA 3.0 evaluates its action in real space to reproducing the exact matrix-vector product in Fourier/Spherical Harmonic space \citep{Pen2000}. 
Second, this equivalence enables a paradigm shift from explicit matrix products to functional operations: rather than being formed or stored explicitly, $\mathbf{H}$ is represented implicitly through its action on a vector $\mathbf{H} \nu = \mathbf{A}^\mathrm{T} (\mathbf{N}^{-1} (\mathbf{A} \nu))$, which is evaluated on demand as a sequence of fast operator calls. 
%Third, the Conjugate Gradient (CG) solver is the natural and mathematically optimal partner to this functional paradigm.
Third, the Conjugate Gradient (CG) solver is naturally suited to this functional paradigm. Because it relies solely on applications of the functional operator $\boldsymbol{\mathcal{A}}^\mathrm{T}(\boldsymbol{\mathcal{N}}^{-1}(\boldsymbol{\mathcal{A}}(\boldsymbol{v})))$ for an arbitrary trial vector, it ideally matches our matrix-free setup. 
Furthermore, CG constructs a sequence of $\boldsymbol{\mathcal{H}}$-orthogonal search directions within the Krylov subspace, leading to substantially faster convergence than steepest descent method. 
% The two methods coincide only in the first iteration.
% Furthermore, unlike steepest descent, with which it coincides only on the first iteration, CG builds a sequence of $\boldsymbol{\mathcal{H}}$-orthogonal search directions in the Krylov subspace, achieving rapid convergence.

This paper is structured as follows. Section~\ref{sec:method} establishes the mathematical foundation of AKRA 3.0, 
%we derive the maximum-likelihood normal equation on the curved sky, reformulate it as a matrix-free operator, and detail the CG solver and preconditioner construction. 
we derive the maximum-likelihood normal equation on the curved sky and reformulate it as a matrix-free linear operator. We then describe the CG solver and the construction of the corresponding preconditioner.
Section~\ref{sec:data_result} presents the application to the DES Y3 \texttt{METACALIBRATION} catalog: we reconstruct convergence maps accounting for realistic mask geometries and spatially varying effective source densities, and using them to measure both non-tomographic and tomographic convergence power spectra.
%then extract non-tomographic and tomographic convergence power spectra directly from the maps.
% non-tomographic is wierd
Section~\ref{sec:conclusion} summarizes our findings and discusses the extension of this methodology to real data from ongoing surveys, as well as its application to future survey programs.
% Section~\ref{sec:conclusion} summarizes our findings and outlines the roadmap for extending this methodology to real data from other ongoing and future surveys.

We adopt the following notational conventions in this paper. Vectors representing physical fields are denoted by italic Greek letters (e.g., $\kappa$, $\gamma$, $\nu$). Matrices are represented by boldface font (e.g., $\mathbf{A}$, $\mathbf{N}^{-1}, \mathbf{H}$), whereas linear operators acting on vectors are distinguished by boldfaced calligraphic font (e.g., $\boldsymbol{\mathcal{A}}$, $\boldsymbol{\mathcal{N}}^{-1}$, $\boldsymbol{\mathcal{H}}$). Finally, the superscripts $^\mathrm{T}$ and $^\dagger$ denote the adjoint (transpose) and Hermitian conjugate of an operation or matrix, respectively.

\section{AKRA 3.0}
\label{sec:method}
Building upon the prior-free, maximum-likelihood foundation established in AKRA~2.0, this section introduces the AKRA~3.0 framework (see Fig.~\ref{fig:workflow}), designed specifically to overcome the computational bottlenecks. 
First, the mask convolution previously performed by explicitly constructing the coupling matrix ${}_{\pm2}\mathbf{M}_{\ell_1 m_1 \ell m}$ in AKRA 2.0, is now evaluated implicitly through a sequence of spin-weighted spherical harmonic transforms (SHT) and diagonal multiplications, reducing the storage from $O(N^2)$ to $O(N)$. 
Second, the direct inversion of the normal-equation is replaced by an iterative (preconditioned) CG solver that requires only the action of $\mathbf{H}$ on a vector instead of its explicit form. Together, these two changes reduce the computational cost from $O(N^3)$ to $O(N_{\mathrm{iter}} \, N^{3/2})$ on full sky, making arcminute-resolution reconstruction over thousands of square degrees practical on a single compute node.

\subsection{AKRA Review} \label{subsec:forward_model}
In ideal case, the relationship between the spin-0 lensing convergence $\kappa$ and the complex observable shear $\gamma$ (defined as $\gamma(\hat{\bm{\theta}}) = \gamma_1 + i\,\gamma_2$) is elegantly established through the curved-sky Kaiser-Squires (KS) framework.
 Because $\gamma$ carries spin weight $+2$, the natural basis for its
harmonic analysis on the sphere is the set of spin-weighted
spherical harmonics ${}_{\pm2}Y_{\ell m}$:
\begin{equation}
    \label{eq:shear_SH}
    \begin{split}
      \gamma(\hat{\bm{\theta}})
        &= \sum_{\ell m} \gamma_{2,\,\ell m}\;
           {}_2Y_{\ell m}(\hat{\bm{\theta}})\,, \\
      \gamma^*(\hat{\bm{\theta}})
           &= \sum_{\ell m} \gamma_{-2,\,\ell m}\;
              {}_{-2}Y_{\ell m}(\hat{\bm{\theta}})\,.
    \end{split}
  \end{equation}
  For a real shear field they are related by
  $\gamma_{-2,\,\ell m} = (-1)^{m}\,\gamma^{*}_{2,\,\ell,-m}$.
  
Analogously to the polarization of the cosmic microwave background (CMB),
the spin-$\pm2$ coefficients are decomposed into parity-even (E-mode)
and parity-odd (B-mode) components~\citep{Shi2025}:
\begin{equation}
    \label{eq:EB_decomp}
    a_{E,\ell m}
      = -\tfrac{1}{2}\bigl(\gamma_{2,\,\ell m}
        + \gamma_{-2,\,\ell m}\bigr)\,,
    \qquad
    a_{B,\ell m}
      = \tfrac{i}{2}\bigl(\gamma_{2,\,\ell m}
        - \gamma_{-2,\,\ell m}\bigr)\,.
  \end{equation}

  Gravitational lensing produces only E-mode shear
  ($a_{B,\ell m}=0$), 
  the shear and convergence harmonics are then related by the
  curved-sky kernel ~\citep{Hu2000}:
  \begin{equation}
    \label{eq:Dl_kernel}
    \gamma_{\pm2,\,\ell m} = -a_{E,\ell m}=D_\ell\;\kappa_{\ell m}\,,
    \qquad
    D_\ell \equiv
      -\sqrt{\frac{(\ell-1)(\ell+2)}{\ell(\ell+1)}}\,.
  \end{equation}
This expression is the spherical generalization of the KS relation.

In a realistic survey, the observed shear is modulated by a survey mask.
Let $M(\hat{\bm\theta})$ denote mask function in real space, the observed (masked) shear field is
\begin{equation}
  \label{eq:masked_shear}
  \gamma^{\rm obs}(\hat{\bm\theta})
    = M(\hat{\bm\theta})\,\gamma(\hat{\bm\theta}),
\end{equation}
where $M(\hat{\bm\theta})=1/0$ for unmasked/masked pixels.

In harmonic space, the pixel-space multiplication by the mask becomes
a convolution that couples modes of different $(\ell,m)$.
As derived in Ref.~\citep{Shi2025}, the masked spin-$\pm2$ harmonic
coefficients are related to the unmasked ones by
\begin{equation}
  \label{eq:mask_coupling}
  \gamma^{M}_{\pm2,\,\ell_1 m_1}
    = \sum_{\ell m}
      {}_{\pm2}\mathbf{M}_{\ell_1 m_1,\,\ell m}\;
      \gamma_{\pm2,\,\ell m}\,,
\end{equation}
where ${}_{\pm2}\mathbf{M}_{\ell_1 m_1,\,\ell m}$ is the
spin-weighted mask transfer matrix of dimension
$(\ell_{\max}+1)^2 \times (\ell_{\max}+1)^2$.
Combining Eqs.~(\ref{eq:EB_decomp}) and~(\ref{eq:mask_coupling}),
the observed masked shear harmonics are linearly related to the true
convergence harmonics.
\begin{equation}
\begin{pmatrix}
  \gamma^{M}_{2,\,\ell_1 m_1} \\
  \gamma^{M}_{-2,\,\ell_1 m_1}
\end{pmatrix}
= 
\begin{pmatrix}
  {}_{2}\mathbf{M}_{\ell_1 m_1,\,\ell m} & 0 \\
  0 & {}_{-2}\mathbf{M}_{\ell_1 m_1,\,\ell m}
\end{pmatrix}
\begin{pmatrix}
   -(a_{E,\ell m} + i\, a_{B,\ell m}) \\
   -(a_{E,\ell m} - i\, a_{B,\ell m})
\end{pmatrix}\,,
\label{eq:mask_linear}
\end{equation}

Setting $a_{B,\ell m}=0$, the masked shear
harmonics become explicitly linear in the convergence harmonics:
\begin{equation}
  \label{eq:mask_linear}
  \begin{pmatrix}
    \gamma^{M}_{2,\,\ell_1 m_1} \\[4pt]
    \gamma^{M}_{-2,\,\ell_1 m_1}
  \end{pmatrix}
  =
  \begin{pmatrix}
    {}_{2}\mathbf{M}_{\ell_1 m_1,\,\ell m} \\[4pt]
    {}_{-2}\mathbf{M}_{\ell_1 m_1,\,\ell m}
  \end{pmatrix}
    (D_\ell\,\kappa_{\ell m})
\end{equation}

We collect the masked spin-$\pm2$ coefficients
$[\gamma^{M}_{2,\,\ell_1 m_1},\;\gamma^{M}_{-2,\,\ell_1 m_1}]$
into a data vector $\mathbf{y}$ of length $2(\ell_{\max}+1)^2$,
the convergence harmonics $\kappa_{\ell m}$ into a signal vector
$\mathbf{x}$ of length $(\ell_{\max}+1)^2$,
and define the forward operator $\mathbf{A}$ that encodes both the
KS relation and the mask coupling.
The resulting linear relation is
\begin{equation}
  \label{eq:linear_model}
  \mathbf{y} = \mathbf{A}\,\mathbf{x} + \mathbf{n},
\end{equation}
where $\mathbf{x}$ is the signal vector of length $(\ell_{\max}+1)^2$, comprising the true convergence harmonics $\kappa_{\ell m}$; $\mathbf{n}$ is the noise vector with covariance
$\mathbf{N}\equiv\langle\mathbf{n}\mathbf{n}^\mathrm{T}\rangle$.

The minimum-variance estimator of $\mathbf{x}$ is obtained by minimizing
$\chi^2 = (\mathbf{y}-\mathbf{A}\mathbf{x})^\mathrm{T}
\mathbf{N}^{-1}(\mathbf{y}-\mathbf{A}\mathbf{x})$.
Setting the gradient to zero yields the normal equations
\citep{Shi2024akra,Shi2025}:
\begin{equation}
  \label{eq:normal_eq}
\underbrace{(\mathbf{A}^{\mathrm{T}}\mathbf{N}^{-1}\mathbf{A})}_{\displaystyle\mathbf{H}}
  \;\hat{\mathbf{x}}
  = \underbrace{\mathbf{A}^{\mathrm{T}}\mathbf{N}^{-1}\mathbf{y}}_{\displaystyle\mathbf{b}}\,.
\end{equation}

with the formal solution
\begin{equation}
  \label{eq:ML_estimator}
  \hat{\mathbf{x}}
  = \mathbf{H}^{-1}\mathbf{b}
  = (\mathbf{A}^{\mathrm{T}}\mathbf{N}^{-1}\mathbf{A})^{-1}
    \mathbf{A}^{\mathrm{T}}\mathbf{N}^{-1}\mathbf{y}\,.
\end{equation}
This is the foundation of the entire AKRA series program~\citep{Shi2024akra,Shi2025,Shi2025HSC}.
The AKRA series can be understood as the natural generalization of the KS idea: rather than inverting the diagonal kernel $D_\ell$ alone, AKRA solves the full inverse problem that accounts for both the
KS relation and the mode coupling induced by the mask, thereby
AKRA successfully restores unbiased mass mapping within realistic survey geometries.
\subsection{From AKRA 2.0 to AKRA 3.0}
\label{subsec:akra2_to_3}
AKRA solves Eq.~(\ref{eq:normal_eq}) to obtain the
unbiased convergence map.
The two generations (AKRA 2.0 and AKRA 3.0) of the algorithm differ fundamentally in how
they evaluate the normal-equation operator $\mathbf{H}$.

\subsubsection{Calculation problem}

In AKRA 2.0~(on full sky, \cite{Shi2025}), the mask transfer matrix
${}_{\pm2}\mathcal{M}_{\ell_1 m_1,\,\ell m}$
(Eq.~\ref{eq:mask_coupling}) is computed explicitly for every pair of
harmonic modes and stored as a dense matrix.
The normal-equation matrix
\begin{equation}
  \label{eq:M_explicit}
  \mathbf{H} \equiv \mathbf{A}^{\mathrm{T}}\mathbf{N}^{-1}\mathbf{A}
\end{equation}
is then assembled and inverted directly (via Cholesky decomposition or
eigendecomposition) to yield
$\hat{\mathbf{x}} = \mathbf{H}^{-1}\mathbf{b}$.

\begin{figure*}[t]
  \centering
  \includegraphics[width=0.95\textwidth]{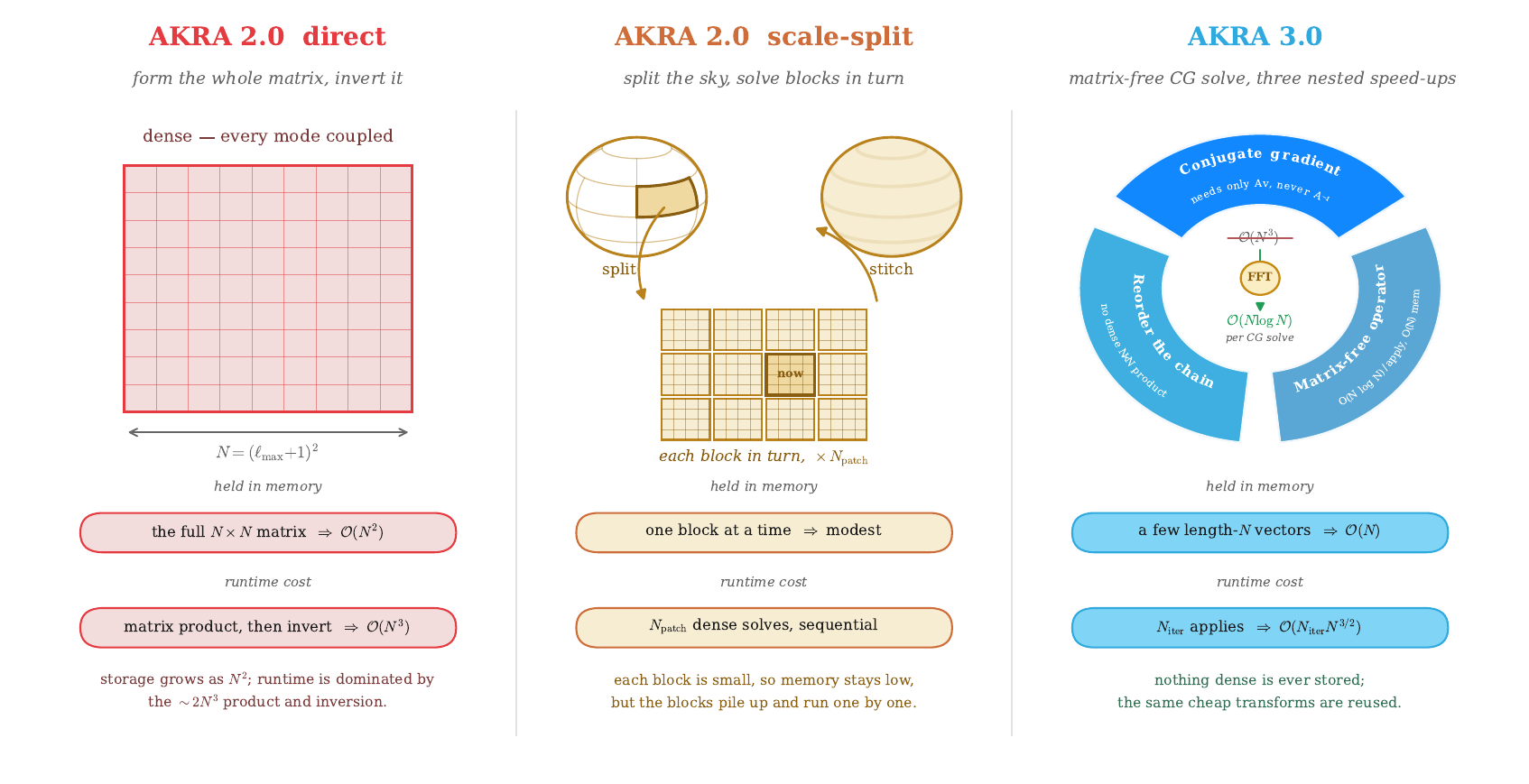}
  \includegraphics[width=0.9\textwidth]{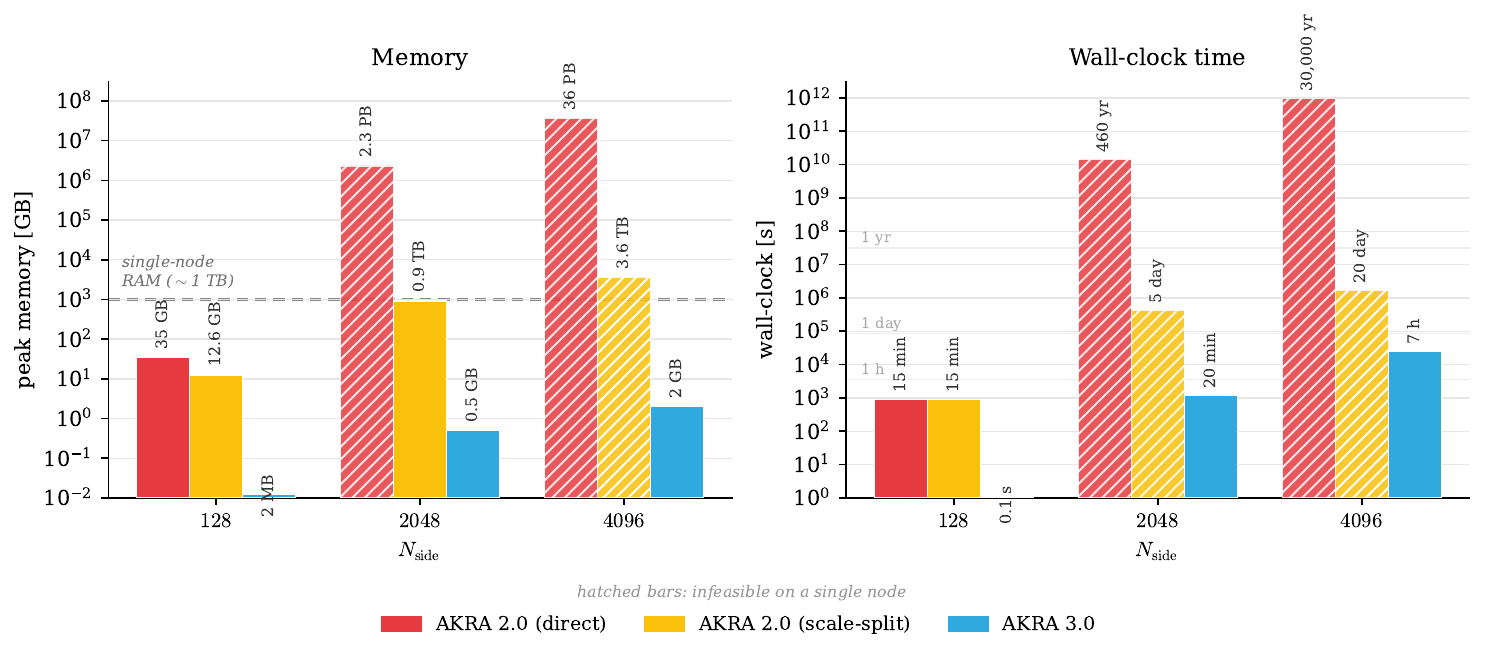}
  \caption{Computational comparison of three methods for solving the normal equations. \textit{Top:} Conceptual diagram illustrating how the normal equations are solved. The approaches differ in memory and speed based on the object held in memory and the operation that dominates runtime: the full dense $N\times N$ matrix (direct), many small dense blocks processed sequentially (scale-split), or a few length-$N$ vectors paired with a reused operator (matrix-free), where $N=(\ellmax+1)^2$.
  \textit{Bottom:} Scaling of peak memory and wall-clock time versus resolution. By reducing the memory footprint by ${\sim}4\times10^{6}$ at $N_{\rm side}=2048$ and ${\sim}2\times10^{7}$ at $N_{\rm side}=4096$, AKRA~3.0 allows the full reconstruction to be run on a single compute node. Specific computational requirements can be estimated using the code available on \href{https://github.com/shiyuan-1/akra_series}{GitHub}.}
  \label{fig:scaling}
\end{figure*}

However, the storage and inversion costs both scale as
high powers of $N=(\ell_{\max}+1)^2$. Specifically, $\mathbf{H}$ contains $N^2$ elements, requiring $\mathcal{O}(N\times N)$
storage, while its direct inversion costs $\mathcal{O}(N^3)$
operations. Figure~\ref{fig:scaling} quantifies the practical consequence of this scaling.  

To circumvent this bottleneck while retaining the AKRA~2.0 formalism
at manageable resolution, AKRA~2.0 introduces a scale-splitting
strategy that decomposes the reconstruction into two complementary
regimes:
\begin{equation}
  \label{eq:scale_split}
  \kappa^{\rm rec}_{\ell m}
    = \kappa^{\rm large}_{\ell m}
    + \kappa^{\rm small}_{\ell m}\,.
\end{equation}
AKRA-sphere addresses this by handling the large scales ($\ell\lesssim\ell_{\rm split}$) on the full sphere using shear maps smoothed with a Gaussian kernel of width $\theta_{\rm sm}$, where the explicit matrix construction remains feasible owing to the modest $\ell_{\max}$ involved.
AKRA-flat handles the small scales ($\ell\gtrsim\ell_{\rm split}$) within flat-sky patches where the flat-sky approximation is valid.
The two reconstructions are combined in harmonic space to produce the final convergence map ~\citep{Shi2025}.

While the scale-splitting strategy extends the practical reach of AKRA~2.0 (Figure~\ref{fig:scaling}, middle column), it does not eliminate the underlying $\mathcal{O}(N^2)$ scaling. The total memory footprint remains substantial at Stage~IV resolutions, and the stitching of patches at the $\ell_{\rm split}$ boundary adds procedural complexity.
AKRA~3.0 overcomes these limitations at their root by replacing
explicit matrix construction with a matrix-free functional operator that operates on the full curved sky at any resolution, as described in the following subsection.

\begin{figure*}[t]
  \centering
  \includegraphics[width=\textwidth]{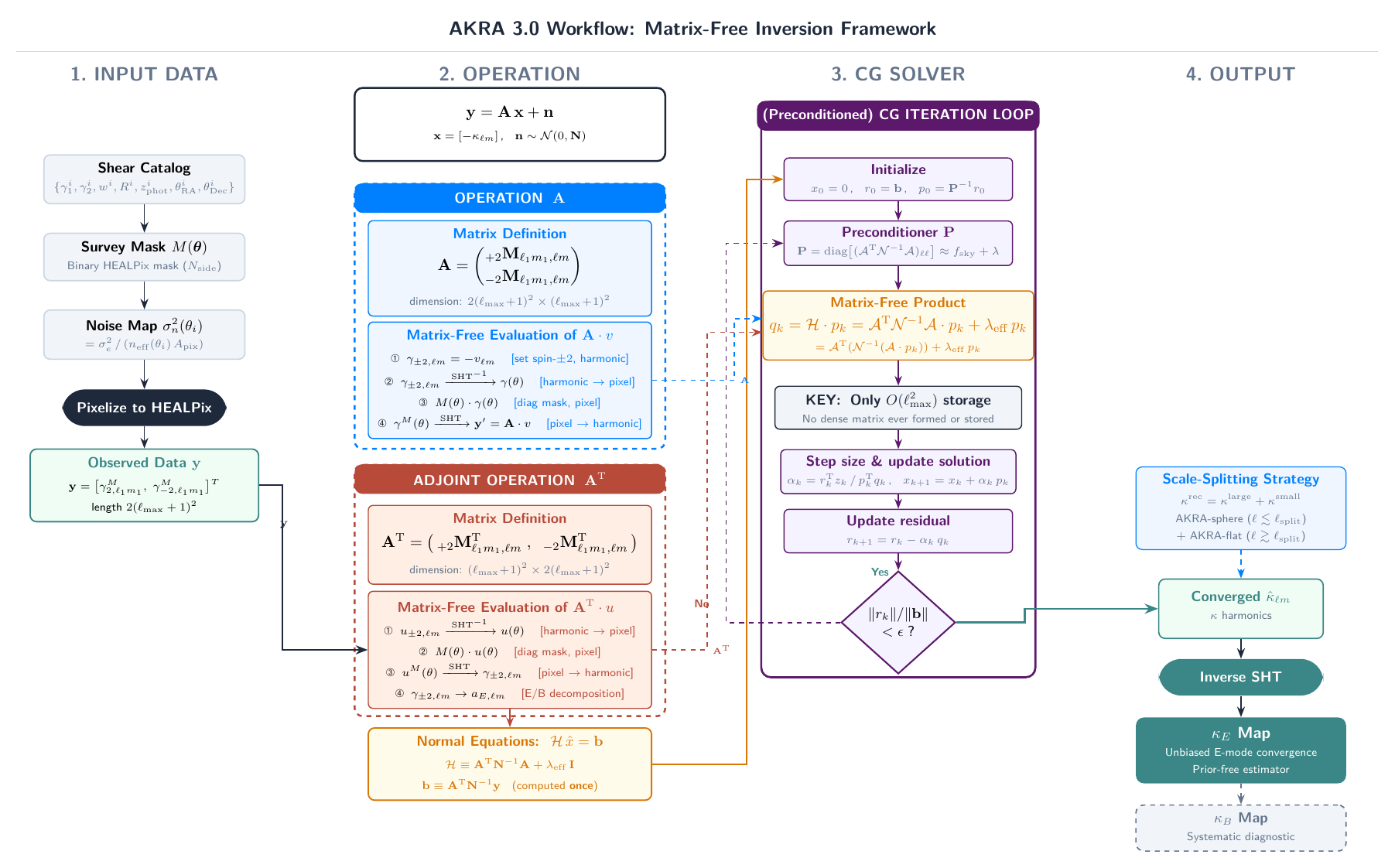}
  \caption{Overview of the AKRA 3.0 matrix-free mass-mapping workflow. Input shear catalogs and survey metadata (column 1) are pixelized into spin-weighted spherical-harmonic data vectors. The operator $\boldsymbol{\mathcal{A}}$ and its adjoint $\boldsymbol{\mathcal{A}}^\mathrm{T}$ (column 2) are never stored as dense matrices but applied implicitly via paired spherical harmonic transforms and masking operations. A (preconditioned) conjugate-gradient solver (column 3) iterates the normal equations $\boldsymbol{\mathcal{H}}\cdot\hat{\boldsymbol{x}} = \boldsymbol{b}$ to convergence. The output (column 4) is an unbiased, prior-free maximum-likelihood convergence map, with $\kappa_B$ serving as a systematics diagnostic. The computational cost of the direct approach (AKRA 2.0) is contrasted with the AKRA 3.0 in Fig. \ref{fig:scaling}, which demonstrates a memory reduction by a factor of ${\sim}10^6$ at $N_{\mathrm{side}}=2048$ alongside the corresponding speed benchmarks.}
  \label{fig:workflow}
  \end{figure*}

\subsubsection{AKRA~3.0: matrix-free Inversion framework}
\label{subsubsec:akra3}
The AKRA 3.0 framework overcomes traditional computational hurdles through a fundamental shift in perspective. It combines three interdependent elements to entirely bypass explicit storage costs and eliminate the need for direct matrix inversion.
% AKRA 3.0 circumvents these computational challenges through a paradigm shift by combining three ingredients that work together:
% : its matrix-free formulation entirely bypasses explicit storage costs, and the CG method eliminates the need for direct inversion.
\begin{enumerate}
  \item \textbf{Conjugate gradient (CG).} Rather than directly inverting the dense matrix $\mathbf{H}$, AKRA 3.0 solves the linear system $\mathbf{H}\hat{\boldsymbol{x}}=\boldsymbol{b}$ iteratively. The key advantage of the CG method is that it never actually requires the matrix $\mathbf{H}$ itself. At each step, it only needs the result of multiplying $\mathbf{H}$ by a trial vector $\boldsymbol{v}$. 
  This changes the core problem from ``how do we store and invert $\mathbf{H}$'' to the much easier ``how do we evaluate $\mathbf{H}\boldsymbol{v}$''.
  \item \textbf{Reordering the operator chain.} Matrix multiplication is inherently associative. Therefore, the product $\mathbf{A}^{\mathrm{T}}\mathbf{N}^{-1}\mathbf{A}\boldsymbol{v}$ can be evaluated from right to left as a sequence of discrete actions on the vector $\boldsymbol{\mathcal{A}}^{\mathrm{T}}\bigl(\boldsymbol{\mathcal{N}}^{-1}(\boldsymbol{\mathcal{A}}(\boldsymbol{v}))\bigr)$ instead of first multiplying the underlying matrices together. We never form the dense product $\mathbf{A}^{\mathrm{T}}\mathbf{N}^{-1}\mathbf{A}$. Instead, each operator smoothly acts on the vector produced by the previous step.
  \item \textbf{A matrix-free operator.} Each action $\boldsymbol{\mathcal{A}}(\boldsymbol{v})$ is performed as a short chain of low-cost steps, a diagonal multiplication, an inverse SHT, another diagonal multiplication, and a forward SHT (\textit{diagonal $\to$ SHT$^{-1}$ $\to$ diagonal $\to$ SHT}), which we detail in the discussion of the storage problem below.
  % In other words, two diagonal multiplications separated by spherical harmonic transforms.
  % This reproduces the dense operator perfectly because of the convolution theorem. 
  Multiplying by the mask matrix in harmonic space is a convolution, and a convolution in one domain is exactly a pointwise product in the conjugate domain. 
  If we write $M(\theta)$ for the mask in pixel space and $\odot$ for the pointwise product, the physical action becomes
    \begin{equation}
      \mathbf{M}\,\boldsymbol{v}
        = \mathrm{SHT}\bigl[\, M(\theta) \odot \mathrm{SHT}^{-1}(\boldsymbol{v}) \,\bigr].
    \end{equation}
    The inverse transform moves the vector to the space where the mask is
    diagonal, the multiplication is a single pointwise product, and the
    forward transform moves the result back. 
    AKRA 3.0 follows a straightforward rule of keeping each operation in the space where it is physically simplest. The mask is simplest in pixel space, so we apply it there. The physical lensing relation is simplest in harmonic space where it just scales each mode by a kernel, so we apply it there. 
    % To translate fields between these two spaces, we use spherical harmonic transforms.
    % The chain therefore yields the
    % same answer as the dense convolution while only ever touching diagonal
    % arrays, which no machine has to store as a $\mathcal{O}(N\times N)$ matrix.
\end{enumerate}

These three pieces interlock. The iterative CG solver simplifies the requirement to just the action $\mathbf{H}\boldsymbol{v}$. Reordering supplies that action as a sequence of operator applications rather than a stored matrix. Factoring the process into transforms and diagonal multiplications scales down the computational complexity of each application dramatically, limited only by the theoretical bound of the underlying spatial transform, such as $\mathcal{O}(N\log N)$ for flat sky FFTs or $\mathcal{O}(N^{3/2})$ for SHTs.
The remainder of this subsection develops the two problems this solves in turn: the
\textbf{storage} problem (how to evaluate $\mathbf{H}\boldsymbol{v}$ without
the matrix) and the \textbf{inversion} problem (how to solve for
$\hat{\boldsymbol{x}}$ without a direct inverse).

\paragraph{\bf The storage problem.}
The central insight of AKRA~3.0 is that iterative solvers do not
require~$\mathbf{H}$ itself, they require only its action $\boldsymbol{\mathcal{H}}$ on a trial vector~$\boldsymbol{v}$. 
The matrix $\mathbf{H}$ is dense because the survey mask couples every harmonic mode to every other. But the physical origin of this coupling is simple: it is a pointwise multiplication by the mask in pixel space. This operation is diagonal, it multiplies each pixel by zero or one. The coupling appears only because AKRA 2.0 translates this pixel-space multiplication into harmonic space, where a pointwise product becomes a dense convolution and represented by the transfer matrix ${}_{\pm2}\mathbf{M}_{\ell_1 m_1,\,\ell m}$ with $O(\ell_{\max}^4)$ elements.

Instead, AKRA~3.0 bypasses this translation. Rather than storing~$\mathbf{A}$ as an explicit matrix, it replaces the matrix-vector product $\mathbf{y}=\mathbf{A}\mathbf{x}$
with a functional evaluation $\mathbf{y}=\boldsymbol{\mathcal{A}}(\boldsymbol{x})$, where the
operator~$\boldsymbol{\mathcal{A}}$ performs exactly the same mapping, from convergence harmonics to observed masked shear, but through a sequence of four operations (\textit{diagonal $\to$ SHT$^{-1}$ $\to$ diagonal $\to$ SHT}) rather than a stored array
of numbers:
\begin{itemize}
  \item Harmonic space (diagonal \footnote{An operation is diagonal when its matrix representation has nonzero values only on the main diagonal. Unlike a general $N\times N$ matrix that mixes all elements of a vector and scales as $\mathcal{O}(N^2)$, a diagonal matrix decouples the system completely. The operation simplifies into an independent pointwise product for each entry, dropping the computational cost dramatically to $\mathcal{O}(N)$.}): multiply the convergence
    harmonics~$\kappa_{\ell m}$ by the KS relation~$D_\ell$
    to obtain the spin-$\pm2$ shear harmonics
    $\gamma_{\pm2,\,\ell m} = -D_\ell\,\kappa_{\ell m}$.
  \item Harmonic $\to$ pixel (SHT$^{-1}$): transform
    $\gamma_{\pm2,\,\ell m}$ to pixel space via an inverse spin-2
    spherical harmonic transform \footnote{We describe the spherical case throughout, as appropriate for full-sky surveys. AKRA~3.0 applies equally to flat-sky analyses by replacing each SHT with a 2D FFT; the four-step structure of $\boldsymbol{\mathcal{A}}$ and the CG solver are otherwise unchanged.} to obtain the shear field~$\gamma(\hat{\boldsymbol{\theta}})$.
  \item Pixel space  (diagonal): multiply by the survey
    mask~$M(\boldsymbol{\theta})$ and the inverse noise
    covariance\footnote{The inverse noise covariance~$\mathbf{N}^{-1}$ is diagonal in
pixel space (for uncorrelated shape noise) and enters the
normal-equation operator~$\mathbf{A}^{\mathrm{T}}\mathbf{N}^{-1}\mathbf{A}$
only once. It may therefore be incorporated equivalently either
(i)~as a single pixel-space multiplication folded into one of the
two operators, here we absorb it into~$\boldsymbol{\mathcal{A}}$,
leaving~$\boldsymbol{\mathcal{A}}^{\mathrm{T}}$ to apply only the
transpose of the mask and harmonic-space operations; or
(ii)~split symmetrically as~$\mathbf{N}^{-1/2}$ applied inside
both~$\boldsymbol{\mathcal{A}}$ and
$\boldsymbol{\mathcal{A}}^{\mathrm{T}}$. Both choices yield
numerically identical matrix--vector products and hence identical
CG iterates.}~$\mathbf{N}^{-1}$, both diagonal in pixel space.
    \item Pixel $\to$ Harmonic (SHT): transform masked shear field to spherical harmonic space.
\end{itemize}

The adjoint~$\boldsymbol{\mathcal{A}}^{\mathrm{T}}$ reverses this
sequence.
Each step is either a diagonal multiplication at
cost~$O(\ell_{\max}^2)$ or an SHT at
cost~$O(\ell_{\max}^3)$, so the action of the full
normal-equation operator on any vector,
\begin{equation}
  \boldsymbol{\mathcal{H}}\cdot\boldsymbol{v}
    = \boldsymbol{\mathcal{A}}^{\mathrm{T}}\!\bigl[\boldsymbol{\mathcal{N}}^{-1}
      \bigl(\boldsymbol{\mathcal{A}}(\boldsymbol{v})\bigr)\bigr],
\end{equation}
is evaluated at only~$O(\ell_{\max}^2)$ storage.
The result is mathematically identical to multiplying by the
explicit matrix~$\mathbf{H}$: the mode coupling is fully captured,
but never written down as a matrix.

\paragraph{\bf The inversion problem.}
With no explicit matrix, direct inversion methods such as Cholesky
decomposition or eigendecomposition are no longer available.
Instead, AKRA~3.0 solves the normal equations iteratively using CG method.
CG refines a trial vector~$\boldsymbol{v}$ toward the solution~$\hat{\boldsymbol{x}}$ by repeatedly applying the operator~$\boldsymbol{\mathcal{H}}$ to the current iterate:
\begin{equation}
  \boldsymbol{\mathcal{H}}\cdot\hat{\boldsymbol{x}} = \boldsymbol{b}, \\ \boldsymbol{b}
    \equiv \boldsymbol{\mathcal{A}}^{\mathrm{T}}(\boldsymbol{\mathcal{N}}^{-1}
      \boldsymbol{x}),
\end{equation}
which is evaluated through the matrix-free $\boldsymbol{\mathcal{A}}$ and $\boldsymbol{\mathcal{A}}^{\mathrm{T}}$ operator sequence described above. Each evaluation costs~$\mathcal{O}(N^{3/2})$, dominated by the SHTs. 
Because~$\boldsymbol{\mathcal{H}}$ is symmetric positive-definite,
CG is guaranteed to converge
\footnote{
The standard convergence bound for preconditioned CG gives, after
$k$ iterations~\citep{Saad2003},
\begin{equation}
  \label{eq:cg_convergence}
  \frac{\|\boldsymbol{r}_k\|}{\|\boldsymbol{r}_0\|}
    \;\leq\; 2\left(
      \frac{\sqrt{\kappa}-1}{\sqrt{\kappa}+1}
    \right)^{\!k},
\end{equation}
where $\kappa \equiv \kappa(\mathbf{P}^{-1}\boldsymbol{\mathcal{H}})$
is the condition number of the preconditioned operator and
$\boldsymbol{r}_k$ is the residual at iteration~$k$.
Requiring the relative residual to fall below a tolerance~$\epsilon$,
the number of iterations scales as
\begin{equation}
  \label{eq:niter}
  N_{\mathrm{iter}}
    \;\sim\; \tfrac{1}{2}\sqrt{\kappa}\;\ln(2/\epsilon).
\end{equation}
}.
For the DES Y3 footprint, a relative residual of $\epsilon = 10^{-3}$ is reached in $N_{\mathrm{iter}}\sim 80$ iterations, giving a total cost of $\mathcal{O}(N_{\mathrm{iter}}N^{3/2})$, which is roughly six orders of magnitude more efficient than the $\mathcal{O}(N^3)$ cost required for direct inversion\footnote{Here $N$ denotes the total number of pixels or modes, with $N\sim\ell_{\max}^2$ on the sphere and $N\sim n^2$ on a flat $n\times n$ grid. The standard separated SHT scales as $\mathcal{O}(N^{3/2})$, while fast algorithms reach $\mathcal{O}(N\log^2\sqrt{N})$ and flat-sky FFTs achieve $\mathcal{O}(N\log N)$. The SHTs can also be GPU-accelerated \citep{Belkner2024}.}.

The iterative solution of normal equations via the (preconditioned)
CG method is well established in CMB data
analysis~\citep{Cantalupo2010,Eriksen2004,Papez2018}, where it is used to handle the large volume of time-ordered data.
The computational context differs from weak lensing. In CMB experiments, the pointing matrix is extremely sparse and presents manageable storage requirements, meaning the CG strategy is motivated by data volume rather than operator density.
Weak lensing presents the exact opposite scenario. The pixelized shear data vector is modest, yet the matrix $\mathbf{H}$ is severely dense in harmonic space due to mask induced mode coupling.
What makes the functional operator approach viable here is the factorization of this dense operator into a \textit{diagonal $\to$ SHT$^{-1}$ $\to$ diagonal $\to$ SHT} sequence, a structure with no direct analogue in CMB mapmaking and not previously exploited for mass mapping.

% We now describe the four stages of the AKRA~3.0 pipeline. Figure~\ref{fig:workflow} provides a schematic overview. In practice, when the survey footprint is small or some harmonic modes 
% are poorly constrained by the data, the normal-equation operator 
% $\boldsymbol{\mathcal{H}}$ can become ill-conditioned.  Following AKRA~2.0~\citep{Shi2025}, we stabilize the inversion by adding a regularization term to $\boldsymbol{\mathcal{H}}$. Because $\boldsymbol{\mathcal{H}}$ carries the overall scale of the inverse noise covariance $\mathbf{N}^{-1}$, a bare identity term $\lambda\mathbf{I}$ would have a physical weight that depends on the units and absolute level of the noise. We therefore scale the regularizer relative to the noise covariance and replace $\lambda\mathbf{I}$ with \begin{equation} \lambda_{\mathrm{eff}}\mathbf{I}, \qquad \lambda_{\mathrm{eff}} \equiv \lambda \big\langle N^{-1}(\boldsymbol\theta)\big\rangle_{M}, \end{equation} where $\langle N^{-1}(\boldsymbol\theta)\rangle_{M}$ denotes the mean of the per-pixel inverse noise variance over the unmasked footprint. We adopt $\lambda = 10^{-3}$ throughout, a value small enough that the regularization has a negligible effect on well-constrained modes while suppressing numerical instabilities in modes that are poorly sampled by the survey geometry. 

\begin{table*}[htbp]
  \centering
  \caption{%
    Summary of the DES~Y3 shear samples analyzed with AKRA~3.0. For each sample we 
    list the photometric redshift range, the HEALPix resolution $N_{\rm side}$ and 
    corresponding pixel area $A_{\rm pix}$ used in the reconstruction, the mean 
    \texttt{METACALIBRATION} shear response $\langle R\rangle$, the residual multiplicative 
    shear-calibration bias $m$ applied as a power-spectrum correction 
    ($C_\ell^{\kappa\kappa} \to C_\ell^{\kappa\kappa}/(1+m)^2$), the sample-averaged 
    effective source density $\bar n_{\rm eff}$, the mean effective galaxy count per 
    pixel $\bar N_{\rm eff} = \bar n_{\rm eff}\,A_{\rm pix}$, the adopted mask 
    threshold $n_{\rm th}$ and its per-pixel equivalent 
    $N_{\rm th} = n_{\rm th}\,A_{\rm pix}$, and the fractional information loss 
    incurred by the mask cut. The two random-split samples are constructed from a 
    galaxy-level random partition of the full shear catalog into two disjoint halves.%
  }
  \label{tab:samples}
  \begin{tabular}{lcccccccccc}
  \hline\hline
  Sample & $z_{\rm phot}$ range & $N_{\rm side}$ & $A_{\rm pix}$ $^{a}$ 
         & $\langle R\rangle$ & $m$ bias$^{b}$
         & $\bar n_{\rm eff}$ & $\bar N_{\rm eff}$ 
         & $n_{\rm th}$ & $N_{\rm th}$ & Info.\ loss$^{c}$ \\
         &                      &                 & [arcmin$^{2}$] 
         &                      & 
         & [arcmin$^{-2}$]      & [pix$^{-1}$]    
         & [arcmin$^{-2}$]      & [pix$^{-1}$]    &             \\
  \hline
  \multicolumn{11}{l}{\textit{Non-tomographic}}\\
  Total          & $0.00$--$1.50$ & $2048$ & $2.95$  & $0.72$ & $-0.020$ & $5.14^{a}$ & $15.16$ & $2.0$ & $5.9$  & $\lesssim 2\%$ \\
  Random split A & $0.00$--$1.50$ & $2048$ & $2.95$  & $0.72$ & $-0.020$ & $2.57$ & $7.58$  & $1.0$ & $2.95$ & $\lesssim 2\%$ \\
  Random split B & $0.00$--$1.50$ & $2048$ & $2.95$  & $0.72$ & $-0.020$ & $2.57$ & $7.58$  & $1.0$ & $2.95$ & $\lesssim 2\%$ \\
  \hline
  \multicolumn{11}{l}{\textit{Tomographic}}\\
  Bin 1 & $0.00$--$0.36$ & $1024$ & $11.80$ & $0.77$ & $-0.006$ & $1.30$ & $15.34$ & $0.5$ & $5.9$ & $\lesssim 2\%$ \\
  Bin 2 & $0.36$--$0.63$ & $1024$ & $11.80$ & $0.73$ & $-0.020$ & $1.30$ & $15.34$ & $0.5$ & $5.9$ & $\lesssim 2\%$ \\
  Bin 3 & $0.63$--$0.87$ & $1024$ & $11.80$ & $0.70$ & $-0.024$ & $1.30$ & $15.34$ & $0.5$ & $5.9$ & $\lesssim 2\%$ \\
  Bin 4 & $0.87$--$2.00$ & $1024$ & $11.80$ & $0.62$ & $-0.037$ & $1.30$ & $15.34$ & $0.5$ & $5.9$ & $\lesssim 2\%$ \\
  \hline\hline
  \end{tabular}
  \vspace{2pt}
  \begin{flushleft}
  \footnotesize
  $^{a}$\,The DES~Y3 \texttt{METACALIBRATION} catalog contains 100{,}204{,}026
  galaxies; after the photometric-redshift cut $0.0 < z_{\rm phot} < 1.5$ the non-tomographic sample retains 100{,}192{,}740 galaxies.  The effective source density $\bar n_{\rm eff}$ depends on the pixelization resolution for the same reason that a measured coastline length depends on the ruler used: at $N_{\rm side} = 4096$ ($A_{\rm pix} \approx 0.74\;\mathrm{arcmin}^2$) the mask retains a slightly different set of pixels than at $N_{\rm side} = 2048$ ($A_{\rm pix} \approx 2.95\;\mathrm{arcmin}^2$), yielding $\bar n_{\rm eff} \approx 5.90\;\mathrm{arcmin}^{-2}$ and $\bar n_{\rm eff} \approx 5.14\;\mathrm{arcmin}^{-2}$, respectively. The specific value does not affect the reconstruction. Once $N_{\rm side}$ is fixed, the mask, the per-pixel noise variance
  $\sigma_n^2(\theta_i)$, and all subsequent pipeline steps are computed
  self-consistently at that resolution.\\[2pt]
  $^{b}$\,For the tomographic bins, $m$ is the central value of the
  Gaussian prior on the per-bin multiplicative shear-calibration bias
  $m^{1}$--$m^{4}$ adopted in the DES~Y3 cosmic-shear analyses
  \citep{Amon2022,Secco2022}. The
  non-tomographic sample uses a single representative value $m=-0.02$.\\
  $^{c}$\,Fractional information loss
  $f_{\rm loss}\equiv
    1-\dfrac{\sum_{i\in\{M=1\}}N_{\rm eff}(\theta_i)}
            {\sum_{i}N_{\rm eff}(\theta_i)}$.
  \end{flushleft}
  \end{table*}

\begin{figure*}[htbp]
  \centering
  \includegraphics[width=0.92\textwidth]{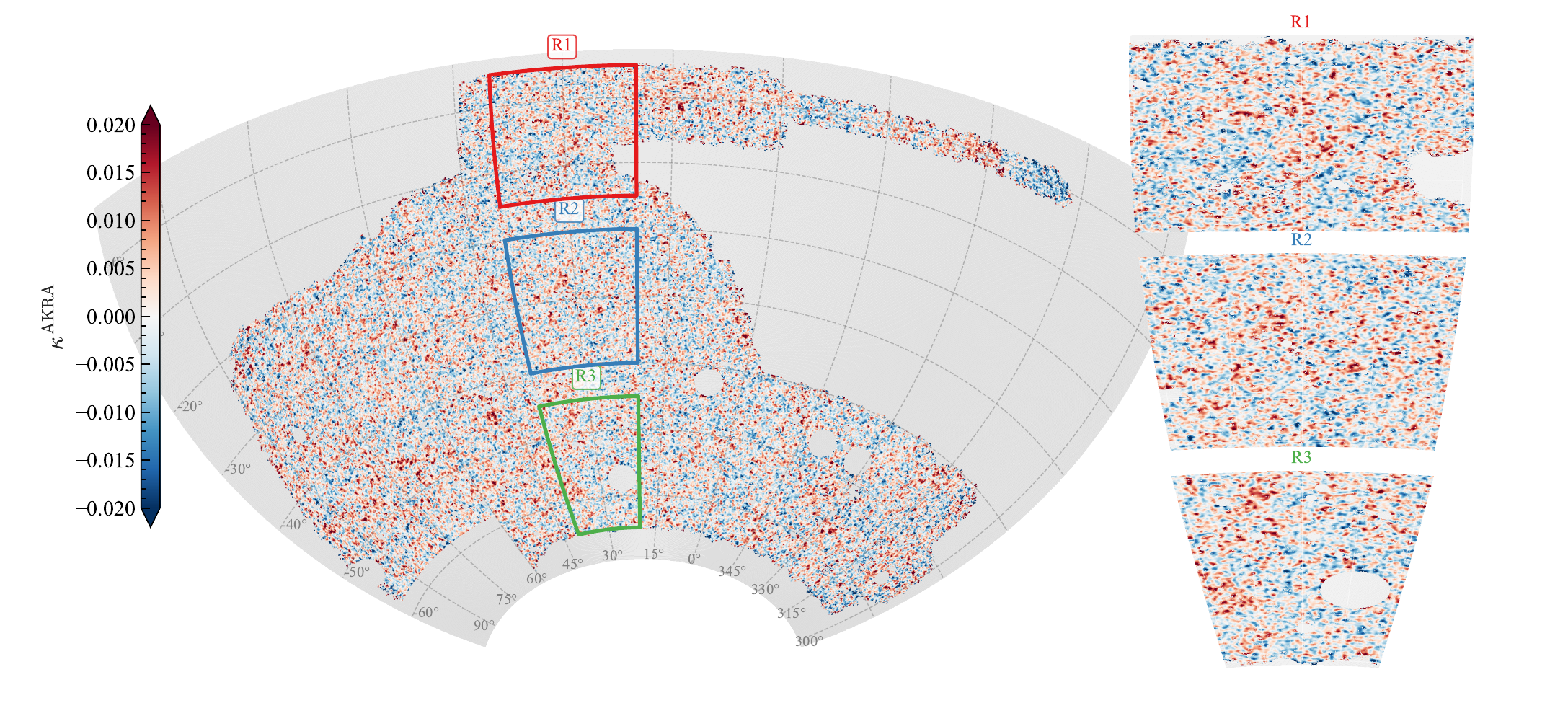}
  \caption{Reconstructed convergence map of the DES~Y3 weak-lensing data obtained 
  with AKRA~3.0. \textit{Left:} Full DES~Y3 footprint, 
  shown in equatorial coordinates. 
  \textit{Right:} Three $10^\circ \times 10^\circ$ zoom-in regions along RA~$=30^\circ$, 
  at Dec~$=-5^\circ$ (R1, top), $-30^\circ$ (R2, middle), and $-55^\circ$ (R3, bottom). The map is produced on a HEALPix grid with $N_{\rm side}=2048$ 
  (pixel scale $\sim$1.72~arcmin) up to $\ell_{\rm max}=4096$, and smoothed with a $6'$ Gaussian kernel for visualization. This represents the highest-resolution, prior-free convergence map of the DES~Y3 footprint produced to date.}
  \label{fig:kappa_nontomo}
\end{figure*}
\vfill

We now describe the four stages of the AKRA~3.0 pipeline. Figure~\ref{fig:workflow} provides a schematic overview. Because the survey is inherently limited to a partial-sky footprint, causing the normal-equation operator $\boldsymbol{\mathcal{H}}$ can become ill-conditioned. Following AKRA~2.0~\citep{Shi2025}, we stabilize the inversion by adding a regularization term $\lambda_{\mathrm{eff}}\mathbf{I}$ to $\boldsymbol{\mathcal{H}}$, the regularized maximum-likelihood estimator (Eq.~\ref{eq:ML_estimator}) then takes the form
% Because $\boldsymbol{\mathcal{H}} = \boldsymbol{\mathcal{A}}^{\mathrm{T}}\boldsymbol{\mathcal{N}}^{-1}\boldsymbol{\mathcal{A}}$ inherits the overall scale of the pixel-space inverse noise covariance, a bare identity term $\lambda\mathbf{I}$ would have a physical weight that depends on the units and absolute level of the noise. 
% We therefore scale the regularizer to the same noise level,
% \begin{equation}
% \lambda_{\mathrm{eff}} \;\equiv\; \lambda\,\big\langle N^{-1}(\boldsymbol\theta)\big\rangle_{M},
% \end{equation}
% where $\langle N^{-1}(\boldsymbol\theta)\rangle_{M}$ is the mean per-pixel inverse noise variance over the unmasked footprint, which reduces to the global mean $\langle N^{-1}\rangle$ in the full-sky limit. The regularized maximum-likelihood estimator (Eq.~\ref{eq:ML_estimator}) then takes the form
\begin{equation}
\hat{\mathbf{x}} \;=\; \big(\mathbf{A}^{\mathrm{T}}\mathbf{N}^{-1}\mathbf{A} + \lambda_{\mathrm{eff}}\,\mathbf{I}\big)^{-1}\,\mathbf{A}^{\mathrm{T}}\mathbf{N}^{-1}\,\mathbf{y},
\label{eq:ML_estimator_reg}
\end{equation}
which AKRA~3.0 evaluates iteratively through the matrix-free CG procedure described above. 
% With this rescaling, $\lambda$ is a dimensionless ratio of the regularizer to the typical pixel-space noise weight, independent of the absolute noise level. 
We adopt a constant regularization parameter of $\lambda = 10^{-3}$ throughout the reconstruction. This threshold ensures that the regularization remains negligible for physical modes well-constrained by the data, while successfully mitigating numerical instabilities.

The convergence rate of the CG method also depends on the condition number of $\boldsymbol{\mathcal{H}}+\lambda_{\mathrm{eff}}\mathbf{I}$, which naturally becomes quite large on a partial sky with spatially varying noise. To accelerate the iteration, we apply a preconditioner $\boldsymbol{\mathcal{P}} \approx (\boldsymbol{\mathcal{H}}+\lambda_{\mathrm{eff}}\mathbf{I})^{-1}$ constructed under the idealized assumption of a completely uniform sky. Incorporating the regularization term yields the exact form
\begin{equation}
  \boldsymbol{\mathcal{P}}
  \;\equiv\;
  \bigl(\langle M^{2}N^{-1}\rangle+\lambda_{\mathrm{eff}}\bigr)^{-1}
  \mathbf{I},
  \label{eq:precond}
\end{equation}
where $\langle\,\cdot\,\rangle$ denotes the pixel average over the
full sphere. Conceptually, AKRA~3.0 thus starts each CG iteration
from a KS-like inverse, then deploys the matrix-free operator $\boldsymbol{\mathcal{H}}$ to adjust that initial guess for the complex realities of mask coupling and inhomogeneity noise. 
Applying $\boldsymbol{\mathcal{P}}$ involves just one scalar multiplication in harmonic space and therefore adds negligible cost to the overall computational loop.

\section{Application to DES Y3 Data}
\label{sec:data_result}
We now apply AKRA~3.0 to the Dark Energy Survey Year~3 (DES~Y3) shear catalog.
We first describe the data preparation and pixelization (Secs.~\ref{sec:data}--\ref{sec:pixelization}), then present the reconstructed convergence maps (Sec.~\ref{sec:map_result}) and the resulting power spectra (Sec.~\ref{sec:ps_result}).
Table~\ref{tab:samples} summarizes the key properties of all samples analyzed in this work, including the photometric redshift ranges, HEALPix resolutions, shear calibration parameters, effective source densities, mask thresholds, and the resulting information loss.

% ----- A. DATA -----
\subsection{Data and shear catalog}
\label{sec:data}

The DES~Y3 public shear catalog~\cite{Gatti2021} contains approximately $10^{8}$ galaxies distributed over $\sim\!4143~\mathrm{deg}^{2}$ in the Southern sky. For each galaxy~$j$, the catalog provides two ellipticity components $\epsilon_1^j$ and $\epsilon_2^j$ measured with the \texttt{METACALIBRATION} algorithm~\cite{Sheldon2017,Gatti2021}, a per-galaxy inverse-variance weight~$w_j$, and photometric redshift information from the \texttt{SOMPZ} scheme~\cite{Myles2021}.

We perform two classes of reconstruction. The baseline is a non-tomographic sample combining all sources with $0.0 < z_{\rm phot} < 1.5$, yielding a single high signal-to-noise convergence map. To exploit redshift information, we additionally split the catalog into the four fiducial DES~Y3 tomographic bins spanning $0.0 < z_{\rm phot} < 2.0$~\cite{Myles2021} and reconstruct each bin independently.

\subsection{Pixelization}
\label{sec:pixelization}
The catalog is binned onto a HEALPix grid at $N_{\rm side}=2048$ ($\sim\!1.72\arcmin$ pixel scale) for the non-tomographic sample, and at $N_{\rm side}=1024$ ($\sim\!3.44\arcmin$) for each tomographic bin. The weight-averaged shear in each pixel is
\begin{equation}
\label{eq:gamma_obs}
    \gamma_{\rm obs}^{\nu}(\theta_{i})
    \;=\;
    \frac{\displaystyle\sum_{j\in\theta_{i}} w_{j}\,\epsilon_{j}^{\nu}}
         {\bar{R}\,\displaystyle\sum_{j\in\theta_{i}} w_{j}},
    \qquad \nu=1,2,
\end{equation}
where $w_{j}$ is the per-galaxy weight and $\bar{R}$ is the mean \texttt{METACALIBRATION} shear response (replaced by a bin-averaged $\langle R\rangle_{b}$ in the tomographic case; see Table~\ref{tab:samples}).

The effective galaxy count per pixel,
\begin{equation}
\label{eq:Neff_pixel}
    N_{\mathrm{eff}}(\theta_i) = \frac{\left(\sum_{j\in\theta_i} w_j\right)^2}{\sum_{j\in\theta_i} w_j^2}\,,
\end{equation}
determines the per-pixel noise variance entering the diagonal inverse-noise operator $\mathbf{N}^{-1}$:
% \footnote{The pixel-level definition of $n_{\mathrm{eff}}(\theta_i) = N_{\mathrm{eff}}(\theta_i)/A_{\mathrm{pix}}$ differs from the conventional survey-averaged quantity $n_{\mathrm{eff}} = A^{-1}(\sum_j w_j)^2/\sum_j w_j^2$; AKRA~3.0 requires the spatially resolved field to populate $\mathbf{N}^{-1}$ in pixel space.}
\begin{equation}
\label{eq:noise_var}
    \sigma_n^2(\theta_i) = \frac{\sigma_e^2}{N_{\mathrm{eff}}(\theta_i)}\,,
\end{equation}
where $\sigma_e \approx 0.26$ is the per-component intrinsic shape dispersion of the DES~Y3 sample.

Pixels with $N_{\rm eff}(\theta_i) < N_{\rm th}$ are masked:
\begin{equation}
  \label{eq:mask}
  M(\theta_i)=
  \begin{cases}
    1, & N_{\rm eff}(\theta_i)\;\ge\;N_{\rm th},\\[2pt]
    0, & \text{otherwise}.
  \end{cases}
\end{equation}
The threshold $N_{\rm th}$ is chosen per sample (Table~\ref{tab:samples}) to exclude pixels whose noise variance would dominate the reconstruction, while keeping the fractional information loss below ${\sim}2\%$.

\begin{figure*}[t]
\centering
  \includegraphics[width=0.45\textwidth]{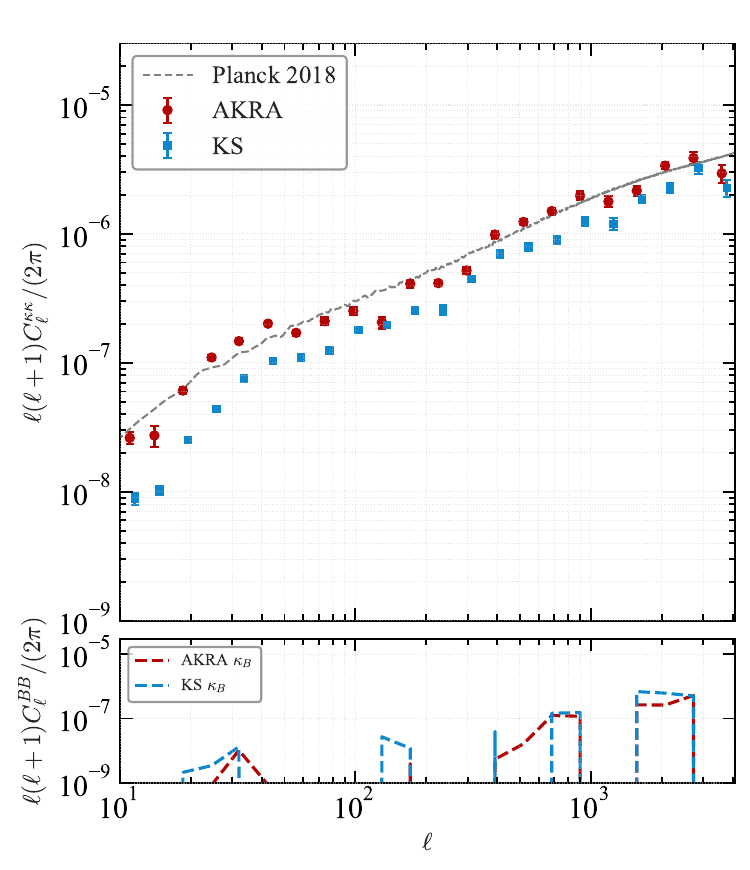}
  \includegraphics[width=0.45\textwidth]{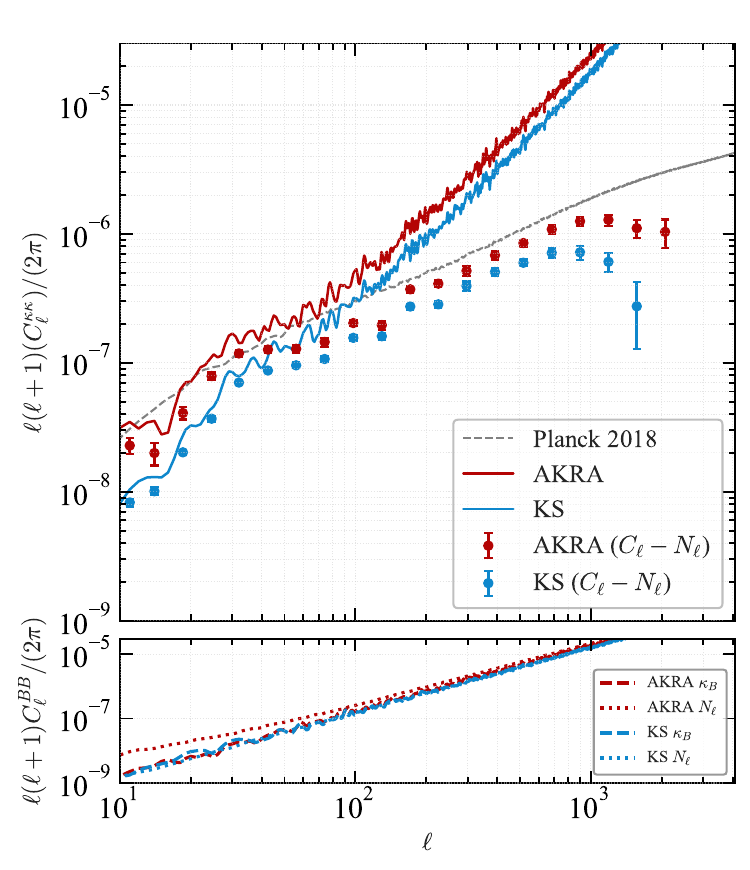}
\caption{%
   Non-tomographic power spectra of the DES~Y3 convergence map reconstructed by AKRA~3.0 (red) and KS (blue), compared with the theoretical $\Lambda$CDM prediction at the fiducial Planck~2018 cosmology (gray dashed curve), shown for reference rather than as a best-fit model.
   \textit{Left:} cross-power spectrum $C_\ell^{\kappa_A\kappa_B}$ between two independent convergence maps reconstructed from disjoint, randomly split sub-samples of the DES~Y3 source catalog. Because the shape-noise realizations of the two sub-samples are statistically independent, the cross-spectrum is free of shape-noise bias at; error bars are estimated from 10 paired noise realizations.
   \textit{Right:} auto-power spectrum $C_\ell^{\kappa\kappa}$ of the map from
   the full sample. The solid lines is the raw spectrum (signal plus shape
   noise) and the points show the spectrum after subtracting the estimated
   shape-noise bias $N_\ell$; error bars are from 50 realizations.
   Noise realizations are built by shuffling the observed shear among the
   unmasked pixels. 
   % The AKRA~3.0 spectrum tracks the fiducial curve, while KS lies systematically below it.
}
\label{fig:cls_nontomo}
\end{figure*}

\begin{figure*}[t!]
  \centering
  \includegraphics[width=0.94\textwidth]{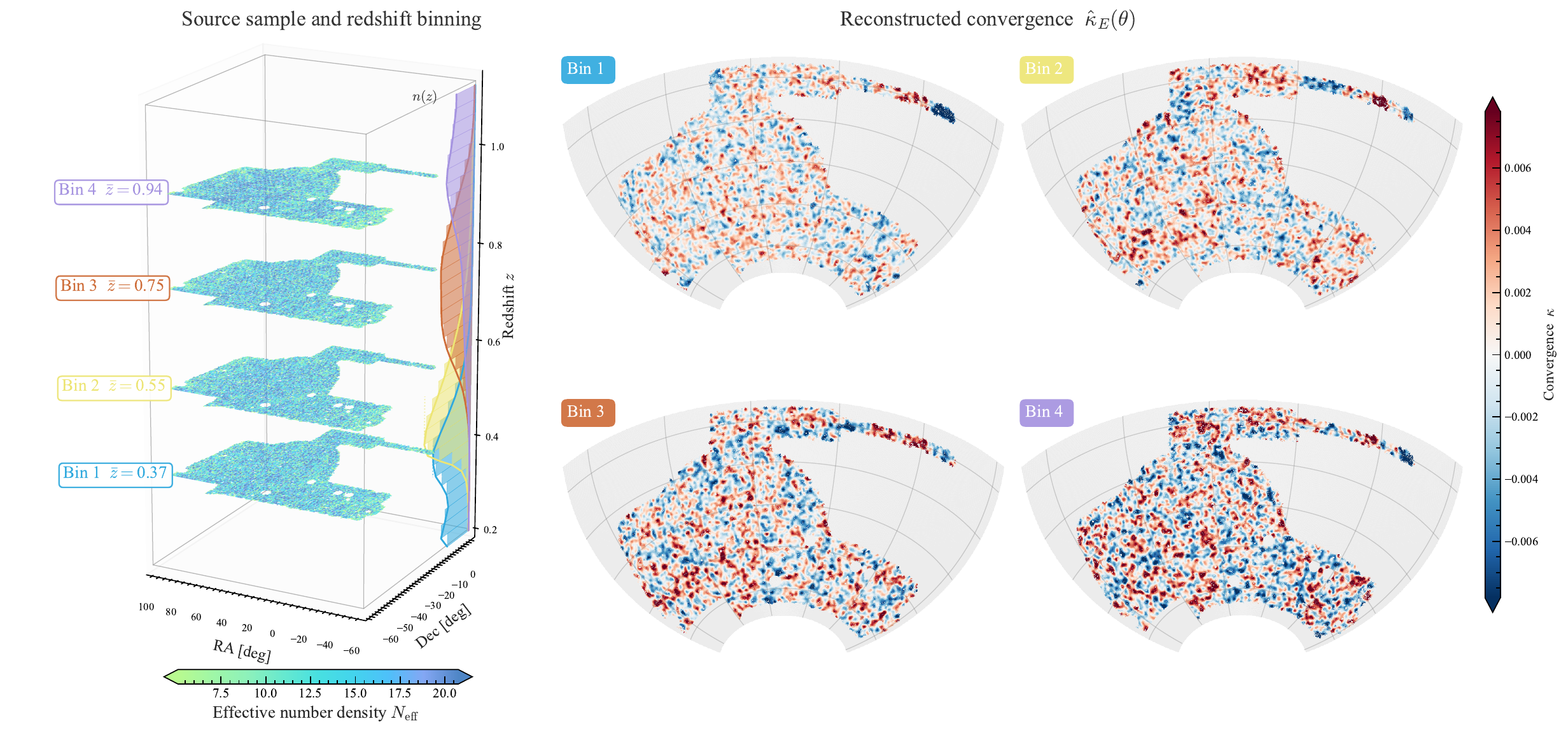}
  \caption{Tomographic AKRA~3.0 reconstruction of the DES~Y3 weak-lensing data 
    across the four source redshift bins. \textit{Left:} Three-dimensional tomographic volume representation. Four effective galaxy number maps $N_{\rm eff}(\theta_i)$  ($N_{\mathrm{side}}=1024$) are displayed as horizontal slices within a box spanned by right ascension (RA), declination (Dec), and photometric redshift $z$, with the vertical axis indicating the mean redshift of each tomographic bin. The corresponding source redshift distributions $n(z)$ are shown to the right, illustrating the radial selection of each bin. 
    \textit{Right:} 
    Reconstructed convergence maps $\kappa_E$ for each bin on the 
    DES~Y3 footprint, shown with a common color scale (red/blue: 
    over-/under-densities of projected matter).The amplitude of $\kappa_E$ increases with source redshift due to the extended lensing path and integrated lensing kernel. Maps are computed at $N_{\mathrm{side}}=1024$ ($A_{\mathrm{pix}}\simeq 11.8\,\mathrm{arcmin}^2$) up to $\ell_{\mathrm{max}}=2048$ and smoothed with a $30'$ Gaussian kernel for visualization. }
  \label{fig:kappa_tomo}
\end{figure*}

% ----- C. MAP RECONSTRUCTION -----
\subsection{Convergence map reconstruction}
\label{sec:map_result}
For all samples, the CG solver is initialized with $\mathbf{x}_0 = \mathbf{0}$ and iterated until the relative residual satisfies $\|r_k\|/\|\mathbf{b}\| < 10^{-3}$, with a regularization parameter $\lambda = 10^{-3}$.
 
\subsubsection{Non-tomographic map}
The reconstruction uses $ N_{\mathrm{side}}=2048, \ell_{\rm max}=4096$ and converges in ${\sim}80$ CG iterations, taking $\sim 0.5$~hour on a single 72-CPU compute node. For comparison, direct inversion at this resolution would be infeasible (Figure~\ref{fig:scaling}).
 
Figure~\ref{fig:kappa_nontomo} shows the resulting $E$-mode convergence map over the full DES~Y3 footprint, with three $10^\circ\times 10^\circ$ zoom panels. This is the highest-resolution, prior-free convergence map of the DES~Y3 data produced to date.
 
\subsubsection{Tomographic maps}
Each of the four tomographic bins is reconstructed independently at $N_{\rm side}=1024$ up to $\ell_{\rm max}=2048$, using the bin-specific mask and effective number density. Figure~\ref{fig:kappa_tomo} displays the four maps alongside the per-bin number density distributions and source redshift distributions $n(z)$. The convergence amplitude increases from the lowest to the highest redshift bin, consistent with the growth of the lensing kernel with source distance.

\subsection{Power spectrum results}
\label{sec:ps_result}
We extract the convergence power spectrum directly from the reconstructed maps and compare it with the standard KS estimator and with the theoretical $\Lambda$CDM prediction at the fiducial \textit{Planck}\,2018 cosmology~\citep{Planck2018}. The fiducial curve is shown for reference only: we do not fit a lensing amplitude or perform cosmological inference here. A quantitative fit would require a reliable covariance that includes
cosmic variance, intrinsic alignments, and marginalization over nuisance parameters such as the multiplicative bias $m$; building this from full pipeline simulations is left to a subsequent work. The goal here is a map-level demonstration that prior-free two-point measurements can be obtained directly from the AKRA~3.0 maps.

The reconstructed map power spectrum inherently contains both the cosmological signal and shape noise. We employ two distinct methods to isolate the physical signal. The clean way is to cross-correlate two maps whose shape-noise realizations are independent, so the cross-spectrum requires no noise model and stands on its own.
The alternative approach auto correlates a single map and subtracts an estimated shape noise bias $N_\ell$. While the auto spectrum achieves higher statistical precision, its accuracy depends entirely on the accuracy of the $N_\ell$ estimate.

\begin{figure*}[h]
  \centering
  \includegraphics[width=0.99\textwidth]{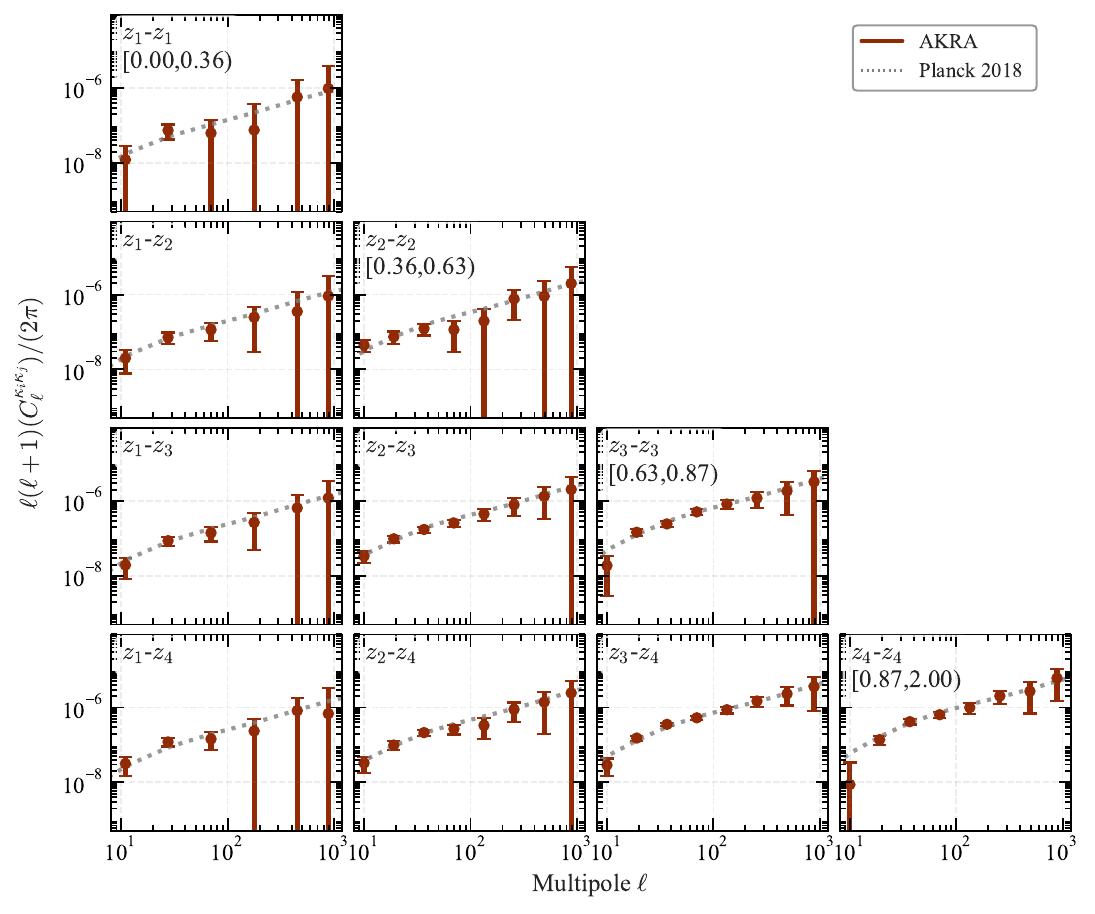}
  \caption{%
    Tomographic auto- and cross-power spectra of the four DES\,Y3
    convergence maps reconstructed by AKRA~3.0 (red points),
    compared with the theoretical $\Lambda$CDM prediction at the
    fiducial Planck~2018 cosmology (gray dashed curves), shown for
    reference rather than as a best-fit model. The $4\times 4$ panel
    matrix displays all ten independent bin pairs $(i,j)$ with
    $1\le i\le j\le 4$: auto-spectra $C_\ell^{\kappa_i\kappa_i}$ on
    the diagonal and cross-spectra $C_\ell^{\kappa_i\kappa_j}$
    off-diagonal. Auto-spectra carry a shape-noise bias $N_\ell^{ii}$
    that is estimated and subtracted with the per-bin pixel-shuffling
    procedure of Sec.~\ref{subsubsec:non-tomo} (error bars from 50
    noise realizations); cross-spectra between distinct bins are
    free of shape-noise bias at the ensemble-mean level because the
    shape-noise realizations of different bins are statistically
    independent.
  }
  \label{fig:cls_tomo}
\end{figure*}

We adopt the cross spectrum as our primary measurement and use the auto spectrum as a supporting check. To estimate $N_\ell$ and compute all error bars, we establish a baseline pixel shuffling procedure. We generate 50 independent noise realizations by randomly shuffling the observed shear values among the unmasked pixels and passing each randomized catalog through the full reconstruction pipeline. This shuffling destroys the spatial correlations of the true cosmological signal while preserving the exact survey geometry and overall noise level, providing an empirical estimate of the background shape noise.
Since the shape noise in independent maps is uncorrelated, this bias vanishes by construction in the cross spectrum, allowing its error estimates to converge rapidly (10 realizations in Figs.~\ref{fig:cls_nontomo} and \ref{fig:cls_tomo}).

\subsubsection{Non-tomographic power spectra}
\label{subsubsec:non-tomo}

Figure~\ref{fig:cls_nontomo} shows the two measurements. For the cross-spectrum (left) we randomly split the source catalog into two disjoint halves, reconstruct each independently and correlate the two maps ($C_\ell^{\kappa_A\kappa_B}$). For the auto-spectrum (right) we reconstruct a single map from the full sample and subtract the shape-noise bias ($C_\ell^{\kappa\kappa}$).

The central result is the cross spectrum. The AKRA 3.0 points follow the fiducial $\Lambda$CDM curve across the measured multipole range, whereas KS lies far below it.
The KS deficit is the expected signature of mask-induced mode coupling, KS leaks power out of the survey footprint, and removing it is precisely what the AKRA forward operator $\mathbf{A}$ is built to do. The mode coupling that motivates the method is
therefore visible directly at the level of $C_\ell$. The auto-spectrum reaches higher signal-to-noise, but after subtracting the shuffling estimate of $N_\ell$ it lies $\sim$10\% below the cross-spectrum. This gap reflects the noise model rather than the map: shuffling the observed shear scrambles the spatial
pattern of $\mathbf{N}^{-1}$ that sets the mask-induced mode coupling, and the shuffled values still carry the signal together with the shape noise, so the estimated $N_\ell$
is biased high and its subtraction removes part of the signal.

The cross-spectrum also sits a few percent below the fiducial
curve\footnote{Although we do not perform cosmological inference here, the offset is conveniently summarized by a single amplitude: scaling the fiducial spectrum by a constant $A_L$ and fitting the binned cross-bandpowers over $10<\ell<2000$ by
inverse-variance-weighted least squares, with errors from the noise realizations, gives $A_L\simeq0.97$. We quote this only as a one-number description of the offset, not as a measurement of the lensing amplitude.}. This is consistent with known
systematics that are not modeled here. The DES Y3 cosmic shear analysis prefers a definitively lower amplitude of matter fluctuations than Planck, with $S_8^{\rm DES Y3}\!\simeq\!0.776$
versus $S_8^{\rm Planck}\!\simeq\!0.832$ \citep{Abbott2022,Planck2018}. If the true convergence field on the DES Y3 footprint reflects this local cosmology, comparing our maps to the fiducial Planck prediction naturally yields a lower amplitude. 
Beyond this intrinsic cosmological difference, the measured power is also influenced by additional unmodeled systematic effects. These include residual multiplicative shear calibration biases \citep{Sheldon2017,MacCrann2020}, photometric redshift errors \citep{Myles2021}, and effects such as intrinsic alignments and baryonic feedback \citep{Chisari2018,vanDaalen2020, ziyang2023}. Because this map level demonstration does not fully model these effects, they introduce additional scatter that can shift the final amplitude.

\subsubsection{Tomographic power spectra}
\label{subsubsec:non-tomo}
 
From the four reconstructed tomographic maps, we extract all ten independent power spectra $C_\ell^{\kappa_i\kappa_j}$ ($1 \le i \le j \le 4$), which are displayed in Fig.~\ref{fig:cls_tomo}. 
% Cross-spectra between distinct redshift bins (the off-diagonal panels) are noise-free by construction at the ensemble-mean level. Conversely, auto-spectra (the diagonal panels) inherently carry a shape-noise bias. We correct for this bias using a pixel-shuffling procedure: for each bin, we generate 50 independent noise realizations by shuffling the observed shear among the unmasked pixels and processing each through the full pipeline. These 50 realizations are used both to subtract the noise bias from the auto-spectra and to estimate the error bars for all measurements. Across all panels, the wide error bars at low multipoles are driven by cosmic (sample) variance; because the DES Y3 footprint covers only roughly 10\% of the sky, each low-$\ell$ bin contains very few independent modes. 

The high-redshift bins give the cleanest test. For $\ell\gtrsim200$ the bin-3 and bin-4 spectra, together with the high signal-to-noise off-diagonal cross-spectra, track the fiducial $\Lambda$CDM curve across the measured multipole range. This strong alignment serves as a robust validation of the tomographic maps, proving them to be rigorous, independent substrates highly capable of supporting future cosmological inference.
The two lowest-redshift bins (bins-1 and bin-2) lie below the fiducial curve near $\ell\sim200$ and carry the largest uncertainties, consistent with the non-tomographic case and with the similar systematics.

Any residual scatter and amplitude deficits remain consistent with the unmodeled systematic effects in the non-tomographic analysis (Sec.~\ref{subsubsec:non-tomo}). In this tomographic case, two specific effects acquire additional bin-dependent structure.  First, the multiplicative bias is corrected per bin using the fixed $m_i$ values listed in Table~\ref{tab:samples}, applied as deterministic point corrections rather than marginalized over a Gaussian prior; thus, residual uncertainties in $m_i$ are not yet propagated into the error budget. Second, photometric-redshift errors $\Delta z_i$ propagate directly into each per-bin lensing kernel. A comprehensive and joint treatment of these systematics alongside the full tomographic covariance is reserved for future work. 

\section{Conclusions and Discussions} \label{sec:conclusion}
The progression from AKRA~2.0 to \akra{} represents a transition from solving the inverse problem by linear algebra to solving it by linear operators.
The maximum-likelihood inverse problem (Eq.~\ref{eq:normal_eq}) was already solved at the level of physics in AKRA~2.0~\citep{Shi2025}; what changes in \akra{} is the recognition that one need not store $\mathbf{H}$ in order to invert it. Recasting the normal equation as
the action of a functional operator
$\boldsymbol{\mathcal{H}} = \boldsymbol{\mathcal{A}}^{\mathrm{T}}
\boldsymbol{\mathcal{N}}^{-1}\boldsymbol{\mathcal{A}}$ and pairing it with a conjugate-gradient solver collapses storage from
$\mathcal{O}(N^2)$ to $\mathcal{O}(N)$ without any prior assumption on the convergence field. This ensures that high-resolution, maximum-likelihood estimation is no longer constrained by the prohibitive $\mathcal{O}(N^2)$ storage and $\mathcal{O}(N^3)$ inversion costs of explicit normal equations.

We apply the \akra{} framework to the DES Y3 \texttt{METACALIBRATION} catalog and reconstruct the matter distribution across the full survey footprint.
Full-footprint reconstruction at $N_{\rm side}=2048$ can be performed on a single compute node without scale splitting.
The convergence power spectra agree with the fiducial $\Lambda$CDM prediction in both the non-tomographic and tomographic configurations, demonstrating that these maps can be used for cosmological inference rather than serving solely as visualizations.
%confirming that these maps are not merely downstream visualizations. 
%They are rigorous, independent substrates capable of future cosmological inference.
% Accordingly, both the non-tomographic and tomographic $\kappa_E$ maps will be released as public data products upon publication.

The power spectrum measured directly from the \akra{} product is the first prior-free, map-level convergence two-point statistics from DES~Y3 data, providing an independent cross-check of the standard shear-based two-point analysis. The \akra{} reconstruction preserves the full statistical information of the shear field while providing direct access to a broad range of map-level statistics, including one-point probability distribution function, peak counts, Minkowski functionals, and wavelet scattering transforms. The map is equally well suited to cross-correlations with other projected fields, including CMB lensing, the Sunyaev--Zel'dovich Compton-$y$ field and X-ray surface brightness, which break parameter degeneracies and self-calibrate systematic
effects across independent probes. 

% As we enter the era of Stage IV surveys such as Euclid, LSST, and CSST, direct inversion is out of reach by orders of magnitude, and AKRA 3.0 is what makes prior-free reconstruction possible. 
As weak-lensing surveys progress toward the Stage IV era, direct inversion becomes out of reach by orders of magnitude, AKRA 3.0 opens a practical route to accurate reconstruction at these scales.
With reconstruction no longer the primary bottleneck, the focus shifts to how convergence maps can be most effectively utilized. Most weak-lensing inference today is built around the shear catalog: theory predictions, covariance models, scale cuts, and likelihood pipelines are all formulated at the level of shear two-point functions. As a direct probe of the projected matter distribution, convergence maps naturally support a broader range of analyses. The challenge ahead is to develop the framework needed to fully exploit this potential. 
% 
% 
%The next step is to build the corresponding infrastructure at the level of the convergence map: analytic and simulation-based covariances for map-level statistics, systematics models that
%propagate from pixel space, and inference frameworks that can combine two-point and non-Gaussian map-level observables in a unified analysis. 

\section*{Acknowledgements}
This work is supported by NSFC Grant No. 12503004, the National Key R\&D Program of China (2023YFA1607800, 2023YFA1607801), and the Fundamental Research Funds for the Central Universities. 
JY acknowledges the support from NSFC Grant No.12573006.
This work made use of the Gravity Supercomputer at the Department of Astronomy, Shanghai Jiao Tong University.

\section*{Data availability statement}
The convergence maps will be made publicly available upon publication. Due to the large size of the dataset, the $B$-mode maps and noise realizations are not included in the public release but will be shared upon reasonable request to the corresponding author.

\appendix

\setcounter{section}{0}
\renewcommand{\thesection}{\Alph{section}}

\section{Derivation of the B-mode}
The signal vector is extended to
$\mathbf{x} = [\kappa^{E}_{\ell m},\,\kappa^{B}_{\ell m}]^{\rm T}$
of length $2(\ell_{\max}+1)^{2}$,
and the forward operator $\mathbf{A}$ maps this joint signal to the
observed masked shear harmonics
$\mathbf{y} = [\gamma^{M}_{2,\ell m},\,\gamma^{M}_{-2,\ell m}]^{\rm T}$.
Retaining both E- and B-mode contributions in
Eq.~(\ref{eq:mask_linear}):
\begin{equation}
  \label{eq:joint_forward}
  \begin{pmatrix}
    \gamma^{M}_{2,\,\ell_1 m_1} \\[4pt]
    \gamma^{M}_{-2,\,\ell_1 m_1}
  \end{pmatrix}
  =
  \begin{pmatrix}
    {}_2\mathbf{M}_{\ell_1 m_1,\,\ell m} & 0 \\[4pt]
    0 & {}_{-2}\mathbf{M}_{\ell_1 m_1,\,\ell m}
  \end{pmatrix}
  \begin{pmatrix}
    -(a_{E,\ell m} + i\,a_{B,\ell m}) \\[4pt]
    -(a_{E,\ell m} - i\,a_{B,\ell m})
  \end{pmatrix}.
\end{equation}

Substituting the relations
$a_{E,\ell m} = D_\ell\,\kappa^{E}_{\ell m}$ and
$a_{B,\ell m} = D_\ell\,\kappa^{B}_{\ell m}$,
the full forward operator takes the block form
\begin{equation}
  \label{eq:A_block}
  \mathbf{A}
  =
  \begin{pmatrix}
    \mathbf{A}_{EE} & \mathbf{A}_{EB} \\[4pt]
    \mathbf{A}_{BE} & \mathbf{A}_{BB}
  \end{pmatrix},
\end{equation}

where
$\mathbf{A}_{EE} = -{}_2\mathbf{M}\,D_\ell$,
$\mathbf{A}_{EB} = -i\,{}_2\mathbf{M}\,D_\ell$,
$\mathbf{A}_{BE} = -{}_{-2}\mathbf{M}\,D_\ell$, and
$\mathbf{A}_{BB} = +i\,{}_{-2}\mathbf{M}\,D_\ell$.
The joint normal equations
$\mathbf{H}\,\hat{\mathbf{x}} = \mathbf{b}$,
with $\mathbf{H} \equiv \mathbf{A}^{\rm T}\mathbf{N}^{-1}\mathbf{A}$,
then expand in $2\times2$ block form as
\begin{equation}
  \label{eq:normal_block}
  \begin{pmatrix}
    \mathbf{H}_{EE} & \mathbf{H}_{EB} \\[4pt]
    \mathbf{H}_{BE} & \mathbf{H}_{BB}
  \end{pmatrix}
  \begin{pmatrix}
    \hat{\kappa}^{E} \\[4pt] \hat{\kappa}^{B}
  \end{pmatrix}
  =
  \begin{pmatrix}
    \mathbf{b}_{E} \\[4pt] \mathbf{b}_{B}
  \end{pmatrix},
\end{equation}
where the four blocks of $\mathbf{H}$ are
\begin{align}
  \mathbf{H}_{EE}
    &= \mathbf{A}_{EE}^{\rm T}\mathbf{N}^{-1}_{+}\mathbf{A}_{EE}
     + \mathbf{A}_{BE}^{\rm T}\mathbf{N}^{-1}_{-}\mathbf{A}_{BE}\,,
  \label{eq:HEE}\\
  \mathbf{H}_{BB}
    &= \mathbf{A}_{EB}^{\rm T}\mathbf{N}^{-1}_{+}\mathbf{A}_{EB}
     + \mathbf{A}_{BB}^{\rm T}\mathbf{N}^{-1}_{-}\mathbf{A}_{BB}\,,
  \label{eq:HBB}\\
  \mathbf{H}_{EB}
    &= \mathbf{H}_{BE}^{\rm T}
     = \mathbf{A}_{EE}^{\rm T}\mathbf{N}^{-1}_{+}\mathbf{A}_{EB}
     + \mathbf{A}_{BE}^{\rm T}\mathbf{N}^{-1}_{-}\mathbf{A}_{BB}\,.
  \label{eq:HEB}
\end{align}
The off-diagonal blocks $\mathbf{H}_{EB}$ and $\mathbf{H}_{BE}$ are
generically nonzero because the mask $M(\hat{\bm{\theta}})$ couples
spin-$+2$ and spin-$-2$ modes of different $(\ell,m)$, mixing parity
under the survey geometry.
They vanish only for a full-sky survey ($M\equiv 1$), in which case
the spin-weighted mask matrices reduce to the identity and
the E- and B-mode sectors decouple exactly.

\paragraph{Conditional B-mode solve.}
Given the converged E-mode solution $\hat{\kappa}^{E}$ obtained from
the E-mode normal equations, the lower block row of
Eq.~(\ref{eq:normal_block}) is
\begin{equation}
  \label{eq:lower_block}
  \mathbf{H}_{BE}\,\hat{\kappa}^{E}
  + \mathbf{H}_{BB}\,\hat{\kappa}^{B}
  = \mathbf{b}_{B}\,.
\end{equation}
Then the B-mode satisfies the reduced normal equation
\begin{equation}
  \label{eq:Bmode_reduced}
  \mathbf{H}_{BB}\,\hat{\kappa}^{B}
  = \underbrace{\mathbf{b}_{B}
    - \mathbf{H}_{BE}\,\hat{\kappa}^{E}}_{\displaystyle\tilde{\mathbf{b}}_{B}}\,.
\end{equation}
The effective right-hand side
$\tilde{\mathbf{b}}_{B} = \mathbf{b}_{B} - \mathbf{H}_{BE}\hat{\kappa}^{E}$
subtracts the E-mode leakage contribution that arises from mask-induced
E--B mixing: the off-diagonal block $\mathbf{H}_{BE}$ transfers power
from the known E-mode signal into the B-mode equation, and must be
removed before solving for $\hat{\kappa}^{B}$.
Omitting this correction would yield B-mode estimates biased by
mask-induced E-to-B leakage, defeating the purpose of the
systematic diagnostic.
The operator $\mathbf{H}_{BB}$ and the correction term
$\mathbf{H}_{BE}\hat{\kappa}^{E}$ are both evaluated matrix-free
through the same sequence of spin-weighted spherical harmonic transforms
and diagonal multiplications described in Sec.~\ref{subsec:akra2_to_3},
at a computational cost of a single forward--adjoint operator pair each.

% The resulting linear system~(\ref{eq:Bmode_reduced}) has dimension
% $(\ell_{\max}+1)^{2}$ rather than $2(\ell_{\max}+1)^{2}$, and is
% solved by a secondary PCG iteration that requires roughly half the
% memory and computation of the primary E-mode solve.
 
Since gravitational lensing produces no intrinsic B-modes to leading
order~\citep{Hu2000}, the converged $\hat{\kappa}^{B}$ serves
exclusively as a systematic diagnostic.

\bibliographystyle{apsrev}
\bibliography{ref}% Produces the bibliography via BibTeX.

\end{document}